\documentclass[aps,pra,reprint,groupedaddress]{revtex4-1}
\usepackage{graphicx}
\usepackage{color}
\graphicspath{{figure/}}
\usepackage{amsmath}
\usepackage{hyperref}
\usepackage{array}
\usepackage{xcolor}
\usepackage{soul}
\usepackage[normalem]{ulem}
\usepackage[all,defaultlines=1]{nowidow}
\graphicspath{{figure/}}
\begin{document}

\title{An atom chip interferometer}

\author{B. Wirtschafter$^{1}$, C. I. Westbrook$^{2}$, M. Dupont-Nivet$^{1}$}
\affiliation{$^{1}$Thales Research and Technology France, 1 av. Augustin Fresnel, 91767 Palaiseau, France\\ 
$^{2}$Universit\'e Paris-Saclay, Institut d’Optique Graduate School, CNRS, Laboratoire Charles Fabry, 91127 Palaiseau, France}

\date{\today}

\begin{abstract}
\noindent
We have realized an interferometer using a thermal cloud of magnetically trapped rubidium-87 atoms on a chip.  
The interferometer resembles a Ramsey interferometer with a state selective spatial splitting of the two internal states as proposed by [M. Ammar, and al., Phys. Rev. A, 91, 053623]. 
The splitting is effected by microwave fields from two on-chip waveguides while the atoms remain magnetically trapped. 
The inferred maximum separation is $1.2\pm 0.1~\mu$m.  
We observe interference fringes with a contrast around 8\% limited by velocity difference of the two interferometer states when we close the interferometer. We develop a model describing this contrast decay.
\end{abstract}

\maketitle

\section{Introduction}

Since its beginnings, the field of atom interferometry has flourished in large part due to many exciting applications.
Accelerometers \cite{Salducci2024}, gravimeters \cite{Abend2016}, gyroscopes \cite{Savoie2018}, and various precision measurement devices have been among the chief applications this field has enabled \cite{Cronin2009,Bongs2019,Geiger2020}.
These devices are typically quite large, but the introduction of atom chips, in which cold atoms are confined by the electromagnetic potentials created by wires deposited on a surface of a few cm$^2$, are an appealing means toward miniaturization.
Their small size implies low power consumption and if the atoms can be kept confined during the entire interferometry sequence, they also promise highly compact devices appropriate for navigation and other applications requiring the transport of the sensor. 
Despite this appeal, chip-based interferometer technology has progressed only slowly. 
There remain many hurdles before the realization of a practical device, but here we demonstrate an interferometer exploiting symmetric microwave dressing to coherently separate and then recombine a cloud of atoms trapped on an atom chip \cite{Ammar2014}. 

There have been a few previous realizations of chip-based atom interferometers \cite{Bohi2009,Schumm2005,Schumm2006,Wang2005,Petrovic2013}.
Using the near-field behavior of microwave fields, microwave potentials can be implemented on an atom chip to apply strong, state-dependent forces on the atoms \cite{Fancher2018} allowing the realization of cold atom interferometers and even to the generation of entanglement \cite{Riedel2010}. 
Particularly interesting was the experiment of reference \cite{Bohi2009} which demonstrated the coherent splitting and recombination of atoms that remained trapped on the chip during the entire interferometer sequence. 
That experiment involved Bose-Einstein condensate (BEC) using state-dependent microwave dressing to spatially separate two different magnetic substates \cite{Agosta1989,Perrin2017}. 
The experiment we describe here is similar but differs in some important respects. 
Instead of using a BEC, we use a thermal gas slightly above the BEC transition. 
This mitigates the effect of collisional shifts in the device. 
We also use a symmetric configuration of coplanar waveguides which should produce equal and opposite shifts in the two trapped states. 
This configuration \cite{Ammar2014} helps to reduce differential phase shifts and reduces dephasing of the two interferometer arms despite our use of a thermal cloud \cite{DupontNivet2014,DupontNivet2017b}. 
It should also permit larger splitting distances.

\begin{figure}
\centering  \includegraphics[width=0.46\textwidth]{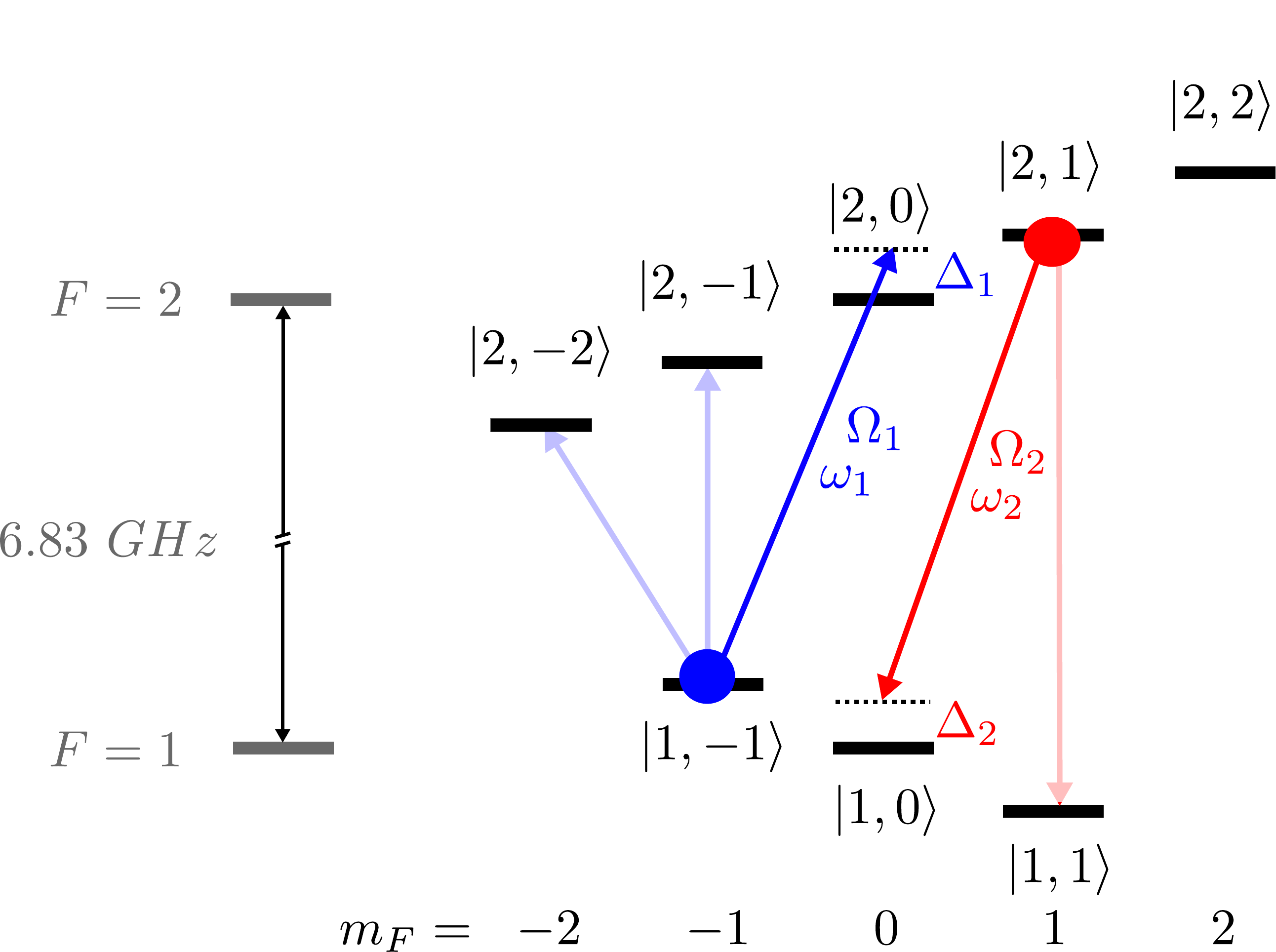}
\caption{\label{fig_RbLevel} (Color online) Levels of $^{87}\mathrm{Rb}$. $5^2S_{1/2}$ levels $\left|1,-1\right>=\left|1\right>$ and $\left|2,1\right>=\left|2\right>$ are the states used for the Ramsey interferometer. 
These two states (displayed in blue and red) are shifted in energy in the presence of microwave fields oscillating at pulsation $\omega_1$ and $\omega_2$, detuned by $\Delta_1$ and $\Delta_2$ from the resonances as shown.
$\Omega_1$ and $\Omega_2$ are Rabi frequencies. 
Transitions in attenuated colors are other possible transitions for microwave dressing.}
\end{figure}

Our interferometer uses an ultra-cold thermal atomic cloud of $^{87}$Rb to perform a modified Ramsey sequence using the states $\left|1,-1\right>=\left|1\right>$ and $\left|2,1\right>=\left|2\right>$ (see figure \ref{fig_RbLevel}). 
These two states can be trapped during the whole interferometry sequence using DC magnetic fields, and selectively split using microwave fields. 
Moreover, around a magnetic field of 3.23~G called the ``magic field", the effective magnetic moments are identical and allow us to minimize the fluctuations of the energy difference between these states \cite{Harber2002,Treutlein2004,Szmuk2015,DupontNivet2025}. 

We report our results on selective and simultaneous displacements of thermal atomic clouds within a magnetic trap using microwave internal state dressing. 
We are able to displace each state by more than 10~$\mu$m.
We also demonstrate interference fringes obtained after a spatial separation of 1.2~$\mu$m.
The fringe contrast is limited by the residual velocity of the atoms after recombination, a problem which can be corrected by an improved separation protocol.
Therefore, we believe that this represents a step towards a useful atom chip interferometer.

The paper is organized as follows: in section~\ref{sec_exp}, we describe the experiment and protocol. Then in section~\ref{sec_depla}, we demonstrate the selective displacement of atomic clouds polarized in either one of the two states of the interferometer, as a function of the microwave frequency of the dressing. 
In section~\ref{sec_splitting}, we show the results obtained when we combine the displacement with a Ramsey sequence, i.e., introducing a spatial splitting and recombination of the two arms of the interferometer before closing the sequence.
In section \ref{sec_conclusion}, we discuss results and insights on how to improve the interferometer phase noise and contrast.
Appendix \ref{Annexe_A_ComputePos} gives a model of the atom motion in the dressed traps while appendix \ref{Annexe_B_Contrast} model the interferometer contrast as a function of the velocity difference between the two interferometer states at the output of the interferometer. 
In appendix \ref{sec_TrapSym}, we compute the traps symmetry parameter.


\section{Experimental protocol}
\label{sec_exp}
\subsection{Atom cooling and preparation}

\begin{figure}
\centering \includegraphics[width=0.48\textwidth]{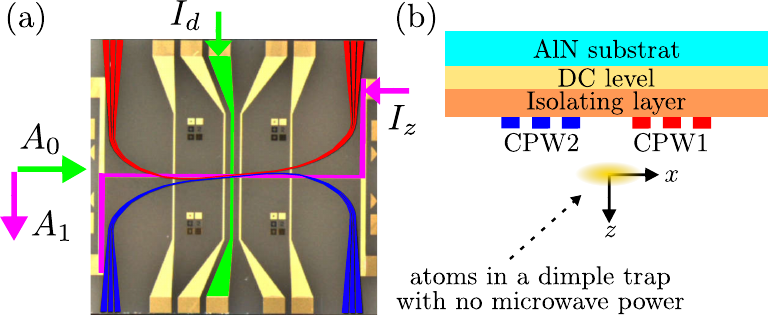}
\caption{\label{fig_AtomChip} (Color online) (a) Picture of the atom chip used in the experiment. 
The two coplanar waveguides (CPW) are highlighted in blue and red. 
The $z$ and dimple wires used in the experiment are highlighted respectively in pink and green, as are the two external magnetic bias fields $A_0$ and $A_1$. 
(b) Vertical slice of the atom chip, at the position of the trap. 
The $z$ and dimple wires are located in the DC level. 
The dimple trap is centered between the two waveguides (CPW1 and CPW2).}
\end{figure}

We use the apparatus described in \cite{Huet2012,Huet2013,DupontNivet2016,Wirtschafter2022}.
About $10^8$ $^{87}$Rb atoms are cooled with a three dimensional magneto-optical trap (3DMOT) using the configuration described in \cite{Farkas2010,Squires2008}, to accomodate the presence of the atom chip. 
This 3DMOT is loaded in approximately 1~s using a cold atom beam created with a two dimensional MOT \cite{Dieckmann1998,Schoser2002}. 
Once the 3DMOT loaded, we turn off the two-dimensional MOT lasers. 
We then allow the atoms to cool further in an optical molasses and then optically pump them into the state $\left|F=2,m_F=2\right>$. 
Next, the atoms are captured in a magnetic trap created by a z-shaped wire and then transferred into a dimple-shaped magnetic trap created by wires of the atom chip and external bias coils (see figure \ref{fig_AtomChip}). 
This transfer protocol is described in \cite{DupontNivet2016,Squires2008,Wirtschafter2022}. 
In the dimple-shaped magnetic trap, the last cooling stage takes place with radio-frequency evaporation. 
After the final cooling, the atomic cloud is transferred to internal state $\left|2,1\right>$ using a microwave stimulated Raman adiabatic passage (STIRAP) process \cite{DupontNivet2015,Vitanov2017}. 
We end up with approximately $10^4$ atoms at a temperature around 800 nK (given for state $\left|2,1\right>$), above the Bose-Einstein condensation threshold which is around 230~nK. 
At this point, atoms are ready for interferometry.

The $\pi/2$ pulses for the interferometer are generated using dedicated microwave and radio-frequency generators to drive the two photons transition between $\left|1,-1\right>$ and $\left|2,1\right>$ \cite{DupontNivet2025}. 
The $\pi/2$ pulse duration is 100~$\mu$s, the radio-frequency is detuned of $-1$~MHz from the transition between $\left|2,1\right>$ and $\left|2,0\right>$ and the frequency of the microwave is tuned to be at resonance or to scan Ramsey fringes.
The microwave is fed into an external horn antenna (which is also used in the STIRAP protocol) directed toward the atoms. 
The radio-frequency is fed into the antenna used for the evaporative cooling.

\subsection{Microwave state selective displacement}

In order to move each state selectively during the interferometer sequence, we create two independent dressed potentials using two microwave fields.  
As shown in figure \ref{fig_RbLevel}, we denote by $\Omega_i$ and $\Delta_i$ the Rabi frequencies and detunings of the fields produced by the two coplanar waveguides. 
These two fields dress the $\left|1,-1\right>$ and the $\left|2,1\right>$ states in such a way that the associated potentials can displace the two states in opposite directions \cite{Ammar2014}.
In the limit $\Omega_i^2/\Delta_i^2 \ll 1$, we can write the potentials as:
\begin{eqnarray}
V_{\left|1\right>}(x) & = & \frac{\mu_B}{2} \Vert\mathbf{B_{dc}}(x)\Vert + \frac{\hbar\Omega_1^2(x)}{4\Delta_1(x)} \;, \nonumber\\
V_{\left|2\right>}(x) & = & \frac{\mu_B}{2} \Vert\mathbf{B_{dc}}(x)\Vert - \frac{\hbar\Omega_2^2(x)}{4\Delta_2(x)} \;,
\label{eq_pot}
\end{eqnarray}
where $x$ is the axis along which the splitting takes place, $\mu_B$ is the Bohr magneton, and $\Vert\mathbf{B_{dc}}(x)\Vert$ is the modulus of the magnetic field of the dimple trap.
The term $\hbar\Omega_i^2(x)/4\Delta_i(x)$ is the AC Zeeman energy shift. 
The spatial dependence of the Rabi frequencies is due to the near field variation of two waveguides (see figure \ref{fig_AtomChip} and appendix \ref{sec_CPWmodel}).

To ensure the microwaves in the waveguides are monochromatic, we use two distinct microwave generators to feed them for the selective displacement of states. 
Signals are amplified through two separate amplifier chains before being injected into the respective waveguide.  
During an experimental run, the frequency and power of the generators remain constant. 
To turn on and off the dressing potentials, we modulate the power to the chip using variable attenuators. 
More details about the setup can be found in \cite{Wirtschafter2022}.

\subsection{Atom detection}
\label{sec_det}

Atoms in both clock states $\left|2,1\right>$ and $\left|1,-1\right>$ are detected by absorption imaging after a short ($<$~1-ms) free expansion during which a magnetic field gradient can be applied to accelerate the fall of the atoms away from the chip. 
This gradient is created with one wire of the atom chip. 
There are two different detection modes depending on whether we want to observe atom displacements or interference fringes.

\textit{Single-detection protocol:} 
To observe atom displacements, we use a single detection pulse and no magnetic field gradient. 
To detect atoms polarized in state $\left|2,1\right>$, we shine a beam on the atoms on the transition between $F=2$ and $F'=3$. 
To detect atoms in state $\left|1,-1\right>$, we first apply a short pulse of MOT light (without repumping beam) that induces spontaneous heating which reduces the atomic density of atoms in $\left|2,1\right>$ to an undetectable level. 
We then use a repumping pulse on the transition between $F=1$ and $F'=2$ to transfer atoms to state $F=2$, followed by a detection pulse on the $F=2$ and $F'=3$ transition. 

\textit{Double-detection protocol:} 
To observe normalized interference fringes, it is necessary to measure both populations. 
After a 500-$\mu$s free expansion during which a short pulse of magnetic field gradient was applied (typically, 170~mA is used in the wire carrying $I_d$, see figure \ref{fig_AtomChip}.a), atoms in state $\left|2,1\right>$ are detected following the \textit{single-detection protocol}. This also moves those atoms out of the depth of field of the imaging system.
Then, still following the \textit{single-detection protocol}, atoms in the state $\left|1,-1\right>$ are detected.
A time interval of 300~$\mu$s is chosen between the two detection pulses so as \cite{DupontNivet2016,Wirtschafter2022} to allow the atoms initially in $F=2$ to separate from those initially in $F=1$ on the absorption image.

\section{Results}
\label{sec_results}

\subsection{State selective displacements}
\label{sec_depla}

\begin{figure*}
\centering  \includegraphics[width=1\textwidth]{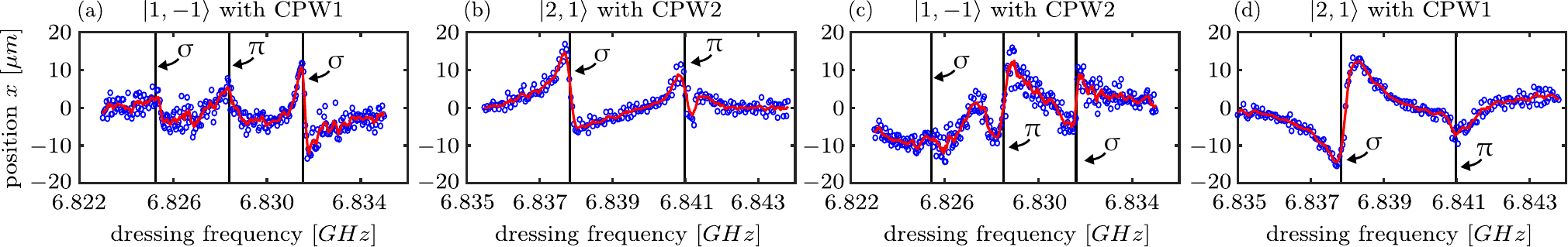}
\caption{\label{fig_deplace} (Color online) Displacements (referred as $x^{cm}_i(t_{m}+t_{t})$ in the text) along $x$ (see figure \ref{fig_AtomChip}) of polarized states as a function of microwave dressing frequency injected into a single coplanar waveguide: 
(a) and (c) displacement of state $\left|1,-1\right>$. 
(b) and (d) displacement of state $\left|2,1\right>$. 
The microwave power is injected in CPW1 ((a) and (d)) or in CPW2 ((b) and (c)), while the power ramp is unchanged. 
Each dot corresponds to one preparation of an atomic cloud and hence a different measurement. 
Solid red lines correspond to a moving average over 5 points. 
Vertical black lines show the position of the allowed transitions calculated with the Breit-Rabi formula for the magnetic field value at the bottom of the trap. 
From state $\left|2,1\right>$ there are 2 possible transitions, while there are 3 transitions for state $\left|1,-1\right>$ (see figure \ref{fig_RbLevel}).}
\end{figure*}
 
For this study, we use the currents and bias fields listed in the second column of table \ref{tab_DCPara}. 
The same table also gives the associated trap parameters.
The resulting magnetic trap has its weakest axis oriented nearly along $x$. 
(The few-degrees tilt between the CPWs and the z wire is negligible for what follows, see figure \ref{fig_AtomChip}.)
The $x$ axis is parallel to the chip and perpendicular to the waveguides and we displace atoms along this axis. 
We do not use the magic field in these measurements in order to obtain larger DC Zeeman splittings between each sublevel, which eases the study of the microwave dressing. 
We scan the microwave dressing frequency around the transitions in order to observe displacements of the atom clouds. 

\begin{table}
\caption{\label{tab_DCPara} Parameters used for creating the DC magnetic trap common to both states $\left|1,-1\right>$ and $\left|2,1\right>$.
$I_d$, $I_z$, $A_0$ and $A_1$ are the trap parameters defined in figure \ref{fig_AtomChip}.
$(\omega_x,\omega_y,\omega_z)/(2\pi)$ are the trap eigenfrequencies computed from the knowledge of $I_d$, $I_z$, $A_0$ and $A_1$.
$B_{bot}$ is the value of the magnetic field at the bottom of the trap, determined by finding the frequency of the RF-knife which completely empties the trap.
The quantity $d_{ct}$ is the measured distance between the chip and the trap center. 
Values are shown for the displacement measurements (section IIIA) and for the interferometer (section IIIB).}
\begin{ruledtabular}
\begin{tabular}{ccc}
 Trap of section: & \ref{sec_depla} & \ref{sec_splitting} \\
 \hline
 $(I_d,I_z)$ & (-280,40)~mA & (-207.2,29.6)~mA \\
 $(A_0,A_1)$ & (6.73,-5.02)~G  & (4.98,-3.71)~G \\
 $(\omega_x,\omega_y,\omega_z)/(2\pi)$ & (164,401,434) Hz & (136,322,350) Hz \\
 $B_{bot}$ & $4.404\pm 0.006$~G & $3.236\pm 0.01$~G \\
 $d_{ct}$ & 28~$\mu$m & 28~$\mu$m
\end{tabular}
\end{ruledtabular}
\end{table}

We begin by preparing a cloud polarized in the $\left|2,1\right>$ or $\left|1,-1\right>$ state in the trap previously described.
At the end of the preparation, we ramp on the microwave field strength in 2~ms at a fixed frequency inside one of the waveguides and then maintain the power constant for 100~$\mu$s. 
We use the maximum microwave power available with our setup which is about 600~mW in the waveguide at the center of the chip. 
DC and microwave fields are then simultaneously turned off in a few hundred $\mu$s (limited by the bandwidth of the power supplies of DC bias coils), hence releasing the atoms. 
Atoms in state $\left|2,1\right>$ ($\left|1,-1\right>$) are detected 100~$\mu$s (400~$\mu$s) later after using the \textit{single detection protocol} described in section \ref{sec_det}. 
We determine the position of the atomic cloud using a 2D-gaussian fit on the optical density and use the position of the maximum of the gaussian fit as the position of the cloud. 
The experiment is repeated for different dressing frequencies, for each state and for each waveguide. 
As the microwave power is ramped on linearly in a time (2~ms) similar to the inverse of the trap frequency along the splitting axis ($(2\pi\times 164)^{-1}\approx 1$~ms), the trap displacement causes the atoms to begin oscillating in the displaced trap. 
Thus, the position of the atomic cloud center of mass is not equal to the position of the trap minimum. 
The center-of-mass position of the cloud in state $\left|i\right>$ in the trap just before the release $x^{cm}_i(t_{m})$ is related to the position of the atomic cloud after the time of flight $x^{cm}_i(t_{m}+t_{t})$ as described in appendix \ref{sec_TOFscale}, where $t_{m}$ is the time from the beginning of the atom displacement and $t_{t}$ is the time-of-flight duration.
In figure \ref{fig_deplace}, we plot the observed displacements of clouds, $x^{cm}_i(t_{m}+t_{t})$ as a function of the microwave frequency in each waveguide. 
The cloud is repelled or attracted depending on the state and on the sign of the detuning.

Since the microwave polarization of the dressing fields are neither perpendicular to the magnetic field of the trap (inducing $\sigma^\pm$ transitions) nor parallel to it (inducing $\pi$ transitions), while scanning the microwave frequency we can induce $\sigma^\pm$ and $\pi$ transitions. As a result, we observe three regions in the spectrum where the state $\left|1,-1\right>$ is displaced (see figures \ref{fig_deplace}.a and \ref{fig_deplace}.c).
The three regions correspond to different microwave polarization components and are shown as the blue arrows in figure~\ref{fig_RbLevel}, as well as being labeled in figure~\ref{fig_deplace}. 
For state $\left|2,1\right>$, only two dressing transitions are possible as can be seen in figure~\ref{fig_RbLevel} and we observe the two corresponding displacements (see figures \ref{fig_deplace}.b and \ref{fig_deplace}.d). 
The direction of each displacement as a function of detuning is determined by the sign of the detuning. 
For state $\left|2,1\right>$, the force is repulsive when $\Delta_2 < 0$ and attractive when $\Delta_2 > 0$, while for state $\left|1,-1\right>$, the force is repulsive when $\Delta_1 > 0$ and attractive when $\Delta_1 < 0$. 
When the microwave frequency is resonant with an atomic transition, the microwave field induces atom losses due to the transfer to untrapped states (only states $\left|1,-1\right>$, $\left|2,1\right>$ and $\left|2,2\right>$ are trappable for our typical magnetic field values).

One expects symmetric displacements for the same state when switching waveguides at fixed frequency (figures \ref{fig_deplace}.a and \ref{fig_deplace}.c, or \ref{fig_deplace}.b and \ref{fig_deplace}.d), but this is not fully the case here. 
There are several possible reasons for this observation. 
(i) Geometrical defects in the waveguides or DC wires are a possibility, but our measurements of the chip wire geometry after fabrication are not consistent with such a large asymmetry. 
(ii) There is a small imbalance of the two microwave fields due to small mismatch between the losses of the waveguides and the microwave components. 
However, independent measurement allowed us to balance the Rabi frequencies, $\Omega_1$ and $\Omega_2$ to within $1\%$, which is insufficient to explain our observations.
(iii) The most likely explanation is that the magnetic trap is not well centered between the two waveguides, which could happen if a bias field perpendicular to the plane of the chip is present \cite{Wirtschafter2022}. 
This would unbalance the force applied to the atoms for a given microwave power between the two configurations.
Despite the observed asymmetry, the data shows that our system allows the selective displacement of atomic clouds in a magnetic trap over distances greater than 10~$\mu$m. 
For example, the 17~$\mu$m observed displacement, $x^{cm}_i(t_{m}+t_{t})$, shown in figure \ref{fig_deplace}.d corresponds to a computed center of mass displacement, $x^{cm}_i(t_{m})$, of 15~$\mu$m (see appendix \ref{Annexe_A_ComputePos}).


\subsection{Interferometry with spatial splitting}
\label{sec_splitting}

\begin{figure}
\centering  \includegraphics[width=0.48\textwidth]{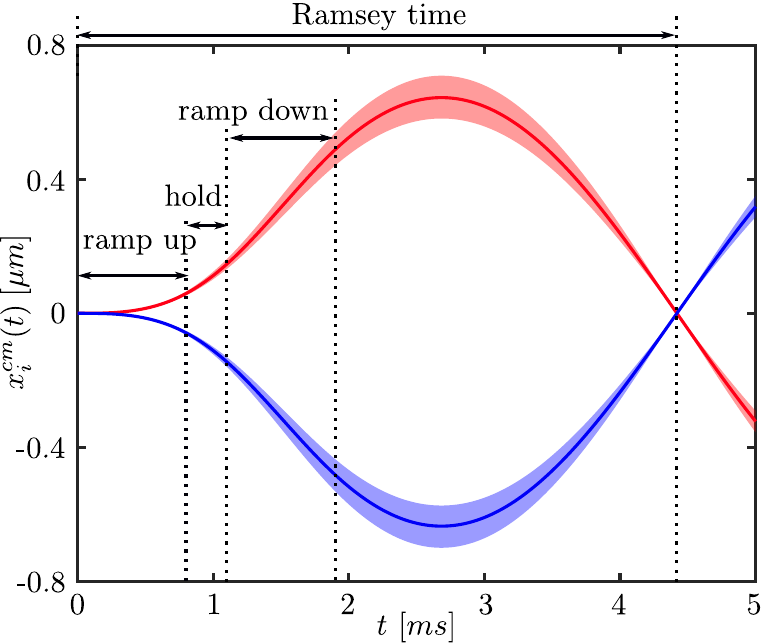}
\caption{\label{fig_Position} (Color online) Simulation of the atom center-of-mass trajectories $x^{cm}_i(t)$ as a function of the time $t$, see appendix \ref{Annexe_A_ComputePos}. 
In red state $\left|2,1\right>$ and in blue state $\left|1,-1\right>$. 
The solid lines are the trajectories and the shaded area represent the uncertainty computed from a 5\% uncertainty on the maximum $\Omega_{i0}$ of the dressing Rabi frequency (see text).}
\end{figure}

\begin{figure}
\centering  \includegraphics[width=0.48\textwidth]{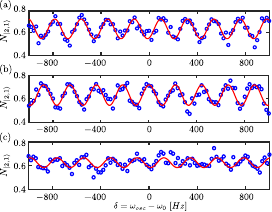}
\caption{\label{fig_Fringes} (Color online) Interference fringes. 
Population in state $\left|2,1\right>$, $N_{\left|2,1\right>}$, as a function of the detuning $\delta$ between the $\pi/2$ pulse frequency $\omega_{osc}$ and the transition frequency $\omega_0$ of $\left|1,-1\right> \leftrightarrow \left|2,1\right>$, with a spatial splitting of the two clock states during the interferometry sequences. 
Blue dots are experimental data and red solid lines are fits.
(a) Microwave injected in CPW1 to displace only $\left|1,-1\right>$. 
(b) Microwave injected in CPW2 to displace only $\left|2,1\right>$. 
(c) Microwaves injected in both waveguides to displace both states simultaneously in opposite directions.}
\end{figure}

To realize an interferometer, a microwave field oscillating at $\omega_1$ is sent to CPW1 to dress $\left|1,-1\right>$, while one at $\omega_2$ is sent in CPW2 to dress $\left|2,1\right>$. 
The interferometry sequence takes place in a different dimple magnetic trap from the displacement measurements since we must work near the magic field.
For the magnetic trap we use the currents and bias fields given in the third column of table \ref{tab_DCPara}. 
This results in a field at the bottom of the trap $B_{bot}$ close to the magic magnetic field (see table \ref{tab_DCPara}). 
The $x$ axis has the same orientation as in the displacement measurements (section \ref{sec_depla}).
The interferometer begins with a $\pi/2$ pulse which drives the two photon transition between the $\left|1,-1\right>$ and $\left|2,1\right>$ states, thus putting them in a coherent superposition. 
We separate the two states with a 800-$\mu$s linear ramp of the microwave dressing powers (fields of figure \ref{fig_RbLevel}). 
Thus the microwave ramps induce oscillations of the atom clouds in the displaced traps. 
Our interferometer protocol thus consists of the 800-$\mu$s splitting ramp followed by a 300-$\mu$s hold time and another 800-$\mu$s ramp down. 
After this sequence, the atoms are oscillating in the trap and we must wait an additional 2.5~ms to allow the two clouds to spatially overlap before applying the second $\pi/2$ pulse \cite{Wirtschafter2022}. 
This leads to a Ramsey time of 4.4~ms.
The parameters used for the two microwave dressing fields are given in table \ref{tab_DressPara}.

\begin{table}
\caption{\label{tab_DressPara} Parameters used for creating repulsive potentials for states $\left|1,-1\right>$ and $\left|2,1\right>$. 
For $\left|1,-1\right>$, $\omega_1$ is sent in CPW1, for $\left|2,1\right>$, $\omega_2$ is sent in CPW2. 
The microwave powers are given at the maximum of the ramp (i.e. during the hold time), the given Rabi frequencies correspond to those maximum of microwave powers.}
\begin{ruledtabular}
\begin{tabular}{ccc}
 Trap for state & $\left|1,-1\right>$ & $\left|2,1\right>$ \\
 \hline
 Dressed transition & $\left|1,-1\right>$ $\leftrightarrow$ $\left|2,0\right>$ & $\left|1,0\right>$ $\leftrightarrow$ $\left|2,1\right>$ \\
 Detuning $\Delta_i/(2\pi)$ & $+$473~kHz & $-$464~kHz \\
 Microwave power & 52.8~mW  & 6.7~mW \\
 Rabi frequency $\Omega_{i0}/(2\pi)$ & 44.6~kHz & 44.9~kHz \\
\end{tabular}
\end{ruledtabular}
\end{table}

Because, unlike in section \ref{sec_depla}, we have nulled the magnetic field perpendicular to the atom chip, we assume that the trap is centered between the CPWs. 
We also assume that the trap is harmonic.
Therefore, in order to maintain the trap symmetry \cite{DupontNivet2014} we need to push the atomic states with an equal force in either direction.
Thus, it is necessary to chose the two microwave powers to keep the induced Rabi frequencies during the hold time ($\Omega_{10}$ and $\Omega_{20}$) as close as possible.
This was done by taking into account the coupling strengths of each transition and the physical properties of the waveguides.

The chosen values of $\Omega_{10}$ and $\Omega_{20}$ as well as the timing of the microwave dressing powers result in a calculated center of mass motion $x_i^{cm}$ for state $\left|i\right>$ displayed in figure \ref{fig_Position}.
We deduced the atom position during the Ramsey sequence using the CPW model of appendix \ref{sec_CPWmodel}, the model of cloud center of mass of \ref{sec_TrapMin} and the microwave ramp model of appendix \ref{sec_COMpos}.
As shown in figure \ref{fig_Position}, the states are displaced by about 0.5~$\mu$m in opposite directions. 
We also see that the two states have different velocities at the output of the interferometer (corresponding to $\dot{x}_1^{cm}=+0.51$~mm/s and $\dot{x}_2^{cm}=-0.53$~mm/s).
This means that the two interferometer paths are at least partially distinguishable. 
When the two states are recombined, they will form a standing wave across the cloud, with a spatial period inversely proportional to $\left|\dot{x}_1^{cm}-\dot{x}_2^{cm}\right|$, which we will spatially average with our imaging system because we are interested in the state $\left|a\right>$ and $\left|b\right>$ population.
But this standing wave limits the observed contrast at the output of the interferometer (see appendix \ref{Annexe_B_Contrast}).

After the second $\pi/2$ pulse, we release the atoms and image them using the \textit{double-detection protocol} described in section \ref{sec_det}. 
In figure \ref{fig_Fringes}, we show interference fringes as a function of the detuning of the $\pi/2$ pulses in which we use one or both of the waveguides to spatially separate the two states between the two pulses. 
The fringes show the detected atom number in state $\left|2,1\right>$ normalized to the total number of atoms.
In figure \ref{fig_Fringes}.a, we sent microwave power only into CPW1 to displace only state $\left|1,-1\right>$. 
The fitted fringe contrast is 0.17 and figure \ref{fig_Position} shows a displacement of $\left|1,-1\right>$ over a maximum distance of $\mathrm{max}\left(x_1^{cm}(t)\right)=+0.57\pm0.06$~$\mu$m.
In figure \ref{fig_Fringes}.b, we sent microwave power only into CPW2 to displace only state $\left|2,1\right>$. 
Here the fitted fringe contrast is 0.18 and figure \ref{fig_Position} shows a displacement of $\left|2,1\right>$ over a maximum distance of $\mathrm{max}\left(x_2^{cm}(t)\right)=-0.58\pm0.06$~$\mu$m.
The similarity of these two results confirms that the displacement is indeed nearly symmetric.
Finally in figure \ref{fig_Fringes}.c we show the fringes obtained when both waveguides are used, displacing the atomic states in opposite directions.
The maximum separation is $1.2\pm0.1$~$\mu$m and the relative velocity is 1.04~mm/s. 
In this case the fringe contrast drops to 0.08.

\begin{figure}
\centering  \includegraphics[width=0.48\textwidth]{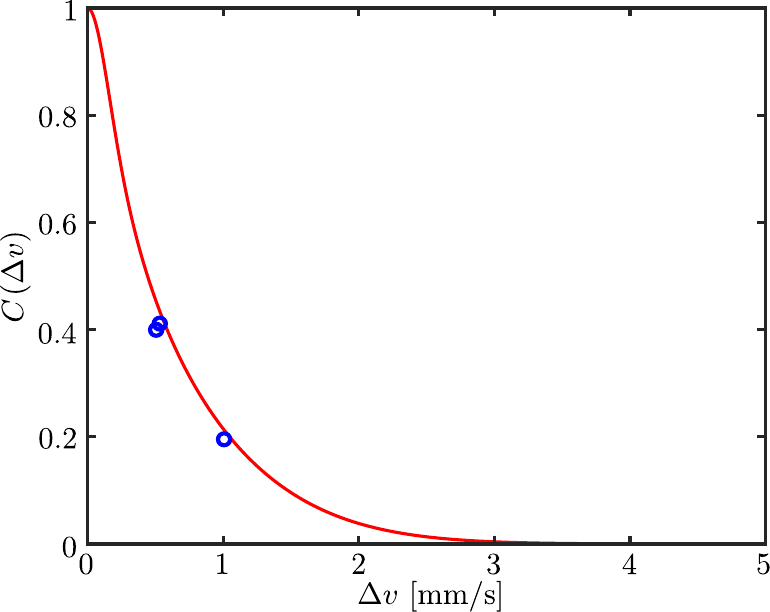}
\caption{\label{fig_ContrastExp} (Color online) Interferometer contrast, $C$, as a function of the velocity difference between the two states at the output of the interferometer $\Delta v$. 
The red solid line is the model described in appendix \ref{Annexe_B_Contrast}, equation (\ref{eq_EqCBE}), with the parameters $N=10^4$ atoms, $T=800$~nK and $\overline{\omega}/(2\pi)=\left( 144\times322\times350 \right)^{1/3}$~Hz.
The blue dots are the experimental data of figure \ref{fig_Fringes}.
We have normalized the observed fringe contrast by the contrast observed without any separation (0.42).
}
\end{figure}

When we apply the same pulse sequence without powering the waveguide, i.e. without any displacement, we find a fringe contrast of 0.42 (limited by, among other things, the imperfect tuning of the $\pi/2$ pulses and the detection pulses, as well as the finite temperature of the atoms).
The loss of contrast in the presence of a spatial displacement can be almost entirely accounted for by the unresolved standing wave pattern due to the different velocities of the two states. 
In appendix \ref{Annexe_B_Contrast}, we describe a model to calculate the contrast, $C_{BE}$, taking this effect into account. 
Figure \ref{fig_ContrastExp} shows the contrast reduction predicted by our model together with the 3 data points described above. 
Surprisingly, despite a cloud temperature which is about 4 times larger than the BEC transition temperature in our conditions, it is necessary to take into account the Bose nature of the particles. 
Bose-Einstein statistics result in a cloud with a somewhat smaller spatial extent, which reduces the  
deleterious effect of the fringes.
It is likely that this effect also accounts for our failure to observe fringes when we attempted to operate the interferometer with larger path separations. 
Larger path separations result in higher relative velocities and therefore in more fringes over the density distribution. 

\begin{figure}
\centering  \includegraphics[width=0.48\textwidth]{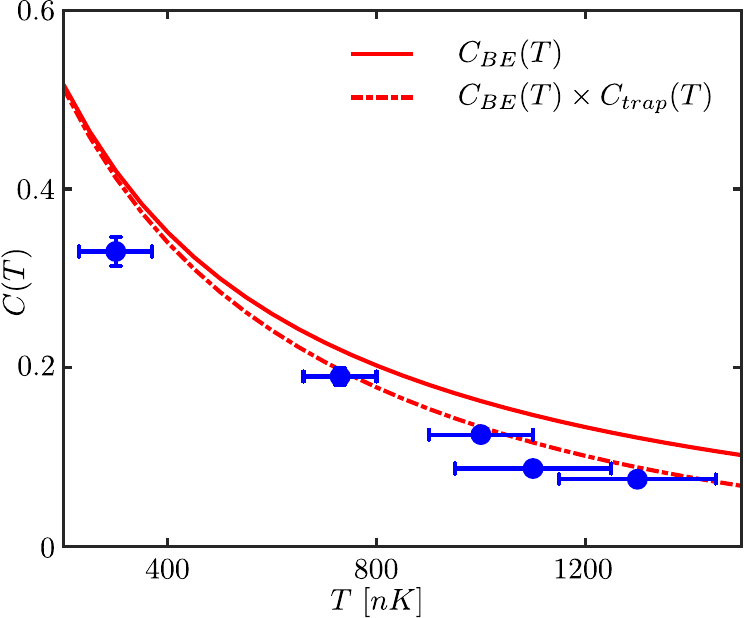}
\caption{\label{fig_ContrastAgainstT} (Color online) Interferometer contrast, $C$, as a function of the atom temperature $T$. 
The blue dots are the experimental data, the two states are displaced in opposite directions with the same parameters as in figure \ref{fig_Fringes}.c.
The temperature was varied by changing the end of the evaporation ramp.
The observed fringe contrast is normalized by the contrast observed without any separation (0.42).
The red solid line is the model described in appendix \ref{Annexe_B_Contrast}, equation (\ref{eq_EqCBE}), for $\Delta v = 1.04$~mm/s. 
The red dashed line shows the additional effect of the estimated trap asymmetry (see appendix \ref{sec_TrapSym}).
At 800 nK the trap symmetry reduces the contrast by 0.87.
}
\end{figure}

The model developed in appendix \ref{Annexe_B_Contrast} also describes the effect of atom temperature. 
A lower temperature results in a narrower position distribution, fewer fringes over that distribution, and thus to an increase of contrast. 
In figure \ref{fig_ContrastAgainstT}, we show measurements of the fringe contrast for different temperatures.
The interferometer configuration is as in figure~\ref{fig_Fringes}.c, that is both waveguides are powered and moved the two states in opposite directions. 
The model which only takes into account the velocity difference (solid lines in figures \ref{fig_ContrastExp} and \ref{fig_ContrastAgainstT}) accounts fairly well for the observations.
However, in appendix \ref{sec_TrapSym}, we give an estimate of the symmetry of the two trapping potentials of $\delta \omega / \omega = 1.5\times 10^{-3}$ and we know that this symmetry leads to a loss of contrast \cite{DupontNivet2016,DupontNivet2017b}.
The effect is small at our temperatures but taking it into account leads to an even better agreement with the observations (dotted line in figure~\ref{fig_ContrastAgainstT}). 

In addition to the loss of contrast, we also observe a small discrepancy between the data points and the fitted curve of figure \ref{fig_Fringes}.
This discrepancy could be due to our detection or to homogeneous dephasing from fluctuations in DC power supplies, residual magnetic fields and microwave amplification. These effects play no role in the variation of the contrast with $\Delta v = \dot{x}^{cm}_2-\dot{x}^{cm}_1$.

\section{Conclusion}
\label{sec_conclusion}

We used thermal clouds with $T~\approx~800~$nK, for which the thermal De Broglie wavelength is $\lambda^T_{DB} \approx 0.2~\mu$m.
Hence, the states were separated by much more than their coherence length $\lambda^T_{DB}$. 
Thus, by the presence of these interference fringes, we demonstrate that we are able to conserve the coherence of our interferometer with a spatial splitting of the two states during the interferometry sequence. 
The fact that the contrast is limited by the velocity mismatch (see appendix \ref{Annexe_B_Contrast}) caused by our splitting procedure indicates that a better adapted pulse sequence, could significantly improve performance, allowing greater path separations.
Among other solutions, recent work \cite{Amri2019,Corgier2018,Ness2018} offers many possibilities for designing improved pulse sequences in which the two arms end up with a near zero velocity mismatch.

In the case of a shot noise limited device, with a contrast $C$, the acceleration sensitivity is given by \cite{DupontNivet2014}:
\begin{equation}
\delta a = \frac{\hbar}{m C \sqrt{N}\int^{T_R}_{0}{\Delta x(t)dt}}
\label{deltaxx}
\end{equation}
where $T_R$ is the Ramsey time, $m$ is the mass of rubidium 87, $\Delta x(t)=\left|x_2^{cm}(t)-x_1^{cm}(t)\right|$ is the distance of separation between the two states of the interferometer as a function of time.
For typical parameters used in this paper, $N~=~10^4$ atoms, $C=10\%$ and $\int^{T_R}_{0}\Delta x(t)dt~\approx~3\cdot 10^{-9}$~m.s (see appendix \ref{Annexe_A_ComputePos} and figure \ref{fig_Position} for computation of this parameter), we found a single-shot sensitivity of $\displaystyle\delta a\approx 23$~mg ($g=9.8$~m.s$^{-2}$). 
Extracting the actual acceleration value from the phase will require studying the interferometer measurement bias, which is related to, among others, cloud kinetic energies and AC Zeeman shifts.

We aim to increase the splitting distance to 100~$\mu$m, the Ramsey time to 40~ms and the contrast to 0.8. 
This would result in single-shot sensitivity in the $\mu$g range.
In order to achieve this performance it will be necessary to improve the indistingushability between the two interferometer paths by reducing the unwanted velocity mismatch $\left| \dot{x}_1^{cm}-\dot{x}_2^{cm} \right|$ and to improve the symmetry of the trapping potentials. 
A reduction of the observed homogeneous phase noise is also needed and a preliminary study \cite{Hello2025c,Hello2025b} has shown that magnetic DC and microwave noise play an important role.

\begin{acknowledgments}
This work has been carried out within the NIARCOS project ANR-18-ASMA-0007-02 funded by the French National Research Agency (ANR) in the frame of its 2018 Astrid Maturation programs. 
This work also received funding from the European Defence Fund (EDF) under grant agreement 101103417 - project ADEQUADE.
Views and opinions expressed are however those of the author(s) only and do not necessarily reflect those of the European Union or the European Commission.
Neither the European Union nor the granting authority can be held responsible for them.
Authors also acknowledge Thales for supporting this research.
\end{acknowledgments}

\appendix


\section{Motion of the atom cloud center of mass}
\label{Annexe_A_ComputePos}

In this appendix, we calculate the position of the center of mass of an atom cloud when the microwave field is turned on in a CPW. 
To achieve this, in section \ref{sec_CPWmodel}, we first model the magnetic field created by a CPW. 
In section \ref{sec_TrapMin}, we then compute the position of the trap potential minimum in the presence of microwave dressing and we solve the differential equation for the position of the center of mass of the atom cloud in section \ref{sec_COMpos}. 
In section \ref{sec_TOFscale}, we also discuss how to scale the trajectories of the atom clouds using the cloud positions measured after a time of flight. 

\subsection{A model for the Rabi couplings created with a CPW}
\label{sec_CPWmodel}

\begin{figure}
\centering  \includegraphics[width=0.35\textwidth]{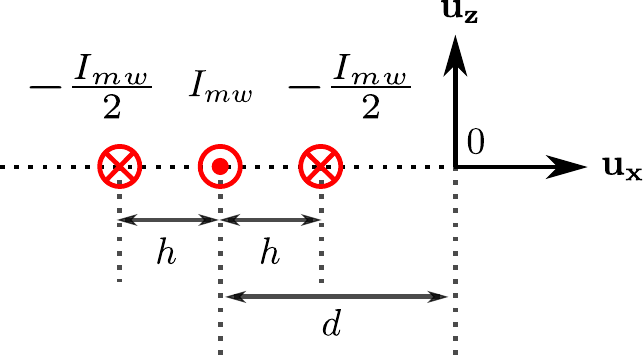}
\caption{\label{fig_CPWgeo} (Color online) Definition of the geometry parameters used to model the magnetic field generated by CPW2 (see figure \ref{fig_AtomChip}), a similar model can be done for CPW1. 
$\mathbf{u_x}$ and $\mathbf{u_z}$ are unit vectors along the $x$ and $z$ axes, respectively. 
Orientation of the axis respective to the atom chip are defined in figure \ref{fig_AtomChip}.}
\end{figure}

As shown in figure \ref{fig_CPWgeo}, a CPW is made of three parallel wires. 
A current $I_{mw}$ flows through the central wire, while a current $-I_{mw}/2$ flows through the two side wires \cite{Wadell1991}.  
The amplitude of the generated magnetic field can be approximated by computing the static field generated by the geometry of figure \ref{fig_CPWgeo} \cite{Ammar2014b}. 
If we consider the three wires as infinitely long and thin, the magnetic field $\mathbf{B_{tot}}$ can be written as:
\begin{widetext}
\begin{eqnarray}
\mathbf{B_{mw}} = \frac{\mu_0 I_{mw}}{2\pi} \left\{ \left[ \frac{x-d}{(x-d)^2+z^2} - \frac{1}{2}\frac{x-d+h}{(x-d+h)^2+z^2} - \frac{1}{2}\frac{x-d-h}{(x-d-h)^2+z^2} \right] \mathbf{u_z} \right. \nonumber\\
+ \left. \left[ - \frac{z}{(x-d)^2+z^2} + \frac{1}{2}\frac{z}{(x-d+h)^2+z^2} + \frac{1}{2}\frac{z}{(x-d-h)^2+z^2} \right] \mathbf{u_x} \right\} \;.
\label{B_microwave}
\end{eqnarray}
The above equation holds for the CPW2 of figure \ref{fig_AtomChip}, and a similar equation holds for CPW1 by replacing $d$ with $-d$. 
This magnetic field can induce $\pi$ and $\sigma^{\pm}$ couplings between levels $\left|F=1,m_{1}\right>$ and $\left|F=2,m_{2}\right>$ (see figure~\ref{fig_RbLevel}). 
The Rabi frequencies $\Omega_{\pi}$ and $\Omega_{\sigma_{\pm}}$ of those couplings are \cite{DupontNivet2016}:
\begin{equation}
\left|\Omega_{\pi}\right|^2 = \left[ \frac{2\mu_B}{\hbar} \left<1,m_1\left|J_z\right|2,m_2\right> \frac{\mathbf{B_{dc}}\cdot\mathbf{B_{mw}}}{\Vert\mathbf{B_{dc}}\Vert} \right]^2 \;,
\quad
\left|\Omega_{\sigma_{\pm}}\right|^2 = \left[ \frac{2\mu_B}{\hbar} \left<1,m_1\left|J_{\pm}\right|2,m_2\right> \frac{1}{2} \left\Vert \mathbf{B_{mw}} - \frac{\mathbf{B_{dc}}\cdot\mathbf{B_{mw}}\mathbf{B_{dc}}}{\Vert\mathbf{B_{dc}}\Vert^2} \right\Vert  \right]^2 \;,
\end{equation}
where $J_\pm$ and $J_z$ are the components of the electronic angular momentum and $\mathbf{B_{dc}}$ is the magnetic field of the dimple trap. 
For the atom chip of figure \ref{fig_AtomChip}, we have $d=50$~$\mu$m and $h=21.5$~$\mu$m, thus we approximate the couplings for $z \ll d$, $h \ll d$ and $x \ll d$ up to the second non-zero order in $x$ as (for CPW2):
\begin{eqnarray}
\Omega_{\pi}(x) = \Omega_{\pi}^0 \left|\frac{z}{d}\right| \left( 1 + \frac{4x}{d} + \frac{5h^2}{3d^2} + \frac{26x^2}{d^2} - \frac{10z^2}{3d^2} \right) \;, \qquad
\Omega_{\sigma_{\pm}}(x) = \Omega_{\sigma_{\pm}}^0 \left( 1 + \frac{3x}{d} + \frac{h^2}{2d^2} + \frac{6x^2}{d^2}\right) \;.
\end{eqnarray}
For CPW1, similar equations hold by replacing $d$ with $-d$. 
\end{widetext}

\subsection{Position of the trap minimum}
\label{sec_TrapMin}

Let us assume that the first term of the potentials of equation (\ref{eq_pot}), the one created by the dimple trap, is harmonic near $x=0$ with a time-independent frequency $\omega$ along the splitting axis $x$. 
As shown in the previous paragraph, the Rabi frequencies of the microwave dressing fields near $x=0$ are (for the $\sigma$ dressing used in section \ref{sec_splitting}):
\begin{equation}
\Omega_i(x,t) \approx \Omega_{i}(t) \left(1+\beta_ix+\alpha_ix^2 \right)
\end{equation}
with $\beta_2=3/(d\gamma)$ for CPW2 dressing the state $\left|2,1\right>=\left|2\right>$, $\beta_1=-3/(d\gamma)$ for CPW1 dressing the state $\left|1,-1\right>=\left|1\right>$, $\alpha_i=6/(d^2\gamma)$ and $\gamma=1+h^2/(2d^2)$. 
The time dependence is due to that of $I_{\mathrm{mw}}$ in equation (\ref{B_microwave}).
This current is ramped up and down to achieve the splitting and merging of the atom clouds. 
This leads to the following time-dependent potentials along the splitting axis:
\begin{equation}
V_{\left|i\right>}(x,t) \approx \frac{1}{2} m \omega^2 x^2 \pm \frac{\hbar\Omega_{i}^2(t)}{4\Delta_i} \left( 1+\beta_ix +\alpha_i x^2\right)^2 \;,
\end{equation}
where the detunings are considered as independent of $x$ and the $+$ (respectively $-$) stand for the state $\left|1,-1\right>$ (respectively $\left|2,1\right>$) trapping potential. 

These potentials can be recast, up to the second order in $x$:
\begin{eqnarray}
V_{\left|i\right>}(x,t) & \approx & \frac{1}{2} m \omega_i^2(t) \left( x - x_{i}(t)  \right)^2 \nonumber\\ 
& & - \frac{1}{2}m\omega_i^2(t)x_{i}^2(t) \pm \frac{\hbar\Omega_{i}^2(t)}{4\Delta_i} \;,
\end{eqnarray}
where $\omega_i(t)$ is the new trap frequency and $x_i(t)$ is the position of the trap minimum:
\begin{eqnarray}
\omega_i^2(t) & = & \omega^2 \pm \frac{\hbar \Omega_{i}^2(t) }{2\Delta_i m} \left( \beta_i^2 + 2\alpha_i \right) \;, \nonumber\\
x_{i}(t) & = & \mp \frac{\hbar \Omega_i^2(t) \beta_i}{2\Delta_i m \omega_i^2(t)} \;.
\label{eq_TrapFreqPos}
\end{eqnarray}
Considering the distance between the two waveguide centers is $2d=100~\mu$m and using the number of section \ref{sec_splitting}, we have $\sqrt{\hbar\Omega_{i0}^2(\beta_i^2+2\alpha_i)/(2|\Delta_i| m)}\approx 2\pi\times$42~Hz. 
Thus the trap frequency along the splitting axis changes from $\omega\approx 2\pi\times$136~Hz with no microwave power to $\omega\approx 2\pi\times$142~Hz when $\Omega_{i}(t)$ is maximum at around 45~kHz.

\subsection{Position of the center of mass of the states $\left|i\right>$}
\label{sec_COMpos}

One can show that the center-of-mass position of the wavefunction of a harmonic oscillator with a time-dependent frequency and position follows the same evolution as a classical one \cite{Lewis1969,Schaff2011}. 
Thus, the center-of-mass position $x^{cm}_i(t)$ of state $\left|i\right>$ is given by:
\begin{equation}
\ddot{x}^{cm}_i + \omega_i^2(t) (x^{cm}_i - x_i(t)) = 0 \;,
\label{eq_Xcm}
\end{equation}
with $\omega_i^2(t)$ and $x_{i}(t)$ given by equations (\ref{eq_TrapFreqPos}).
In the experiment described in section \ref{sec_splitting}, the microwave power is ramped up and down linearly in a time $\tau$ and held constant for a time $\tau_h$, thus we model $\Omega_i^2(t)$ as:
\begin{equation}
\Omega_i^2(t) = \left\{
	\begin{array}{lll}
		\Omega_{i0}^2 t/\tau & \text{if} & 0\leq t < \tau \\
		\Omega_{i0}^2 & \text{if} & \tau \leq t < \tau + \tau_h \\
		\Omega_{i0}^2 (1 - t/\tau) & \text{if} & \tau + \tau_h \leq t < 2\tau + \tau_h \\
		0 & \text{if} & 2\tau + \tau_h \leq t \;.
	\end{array}
\right.
\label{eq_MwPower}
\end{equation}
Using the experimental parameters of section \ref{sec_splitting}: $\tau=800$~$\mu$s, $\tau_h=300$~$\mu$s, and those of table \ref{tab_DressPara}, equation (\ref{eq_Xcm}) allows us to compute the trajectories for the two interferometer states. 
The results are displayed in figure \ref{fig_Position}.  
The observed Ramsey time of 4.4~ms was determined by optimizing the fringe contrast. 
Accordingly, we used the corresponding value of $\omega = 2\pi\times 144$~Hz instead of the simulated one (136~Hz).
The difference is within our estimated uncertainty.

We estimate a 5\% uncertainty on the knowledge of the maximum $\Omega_{i0}$ of the dressing Rabi frequency, which translated into a 5\% uncertainty on $\Omega_i(t)$, and then on the $x_i^{cm}(t)$ uncertainty displayed on figure \ref{fig_Position}. 

Knowing $x^{cm}_i(t)$ allows us to compute the center-of-mass velocity and thus the velocity difference between the two parts of the atom wavefunction when the second $\pi/2$ shines on the atoms. 
At the output of the interferometer, for the data of figure \ref{fig_Fringes}.c, the velocity of the $\left|1\right>$ (respectively $\left|2\right>$) state is $\dot{x}^{cm}_1$ (respectively $\dot{x}^{cm}_2$):
\begin{equation}
\dot{x}^{cm}_1 = + 0.51 \text{ mm/s} \;, \qquad \dot{x}^{cm}_2 = - 0.53 \text{ mm/s} \;.
\end{equation}

\subsection{Scaling the trajectories with the position measured after free expansion}
\label{sec_TOFscale}

After the cloud is released, the center of mass of each cloud continues to evolve. 
In the following, we show how to relate the observed cloud position to its position in the interferometer.
Neglecting the change of the trap frequency with the dressing microwave fields, i.e. $\omega_i(t)\approx\omega$, the equation of motion of the wavefunction center of mass (\ref{eq_Xcm}) is:
\begin{equation}
\ddot{x}_i^{cm} + \omega^2\left(x^{cm}_i-x_i(t)\right) = 0 \;.
\label{eq_A12}
\end{equation}
Let us consider two dimensionless functions $f_i$ and $f_i^{cm}$ such that $x_i(t)=\Delta x_i f_i(t)$, $x_i^{cm}(t)=\Delta x_i f_i^{cm}(t)$. $\Delta x_i$ is the maximum displacement of the trap minimum. We define $f_i(t)$ as:
\begin{equation}
f_i(t) = \left\{
	\begin{array}{lll}
		t/\tau & \text{if} & 0\leq t < \tau \\
		1 & \text{if} & \tau \leq t < \tau + \tau_h \\
		(1 - t/\tau) & \text{if} & \tau + \tau_h \leq t < 2\tau + \tau_h \\
		0 & \text{if} & 2\tau + \tau_h \leq t  \;.
	\end{array}
\right.
\label{eq_MwPower}
\end{equation}
Thus equation (\ref{eq_A12}) becomes:
\begin{equation}
\ddot{f}_i^{cm} + \omega^2\left(f^{cm}_i-f_i(t)\right) = 0 \;.
\end{equation}
If we interrupt the Ramsey sequence at a time $t_{m}$ after the first $\pi/2$ pulse and at the same time the two trapping potentials $V_{\left|1\right>}$ and $V_{\left|2\right>}$ are abruptly switched off, the expansion starts with the following initial conditions (we only consider the splitting axis $x$ which is perpendicular to the gravity) for the position $\Delta x_i f_i^{cm}(t_{m})$, the velocity $\Delta x_i \dot{f}_i^{cm}(t_{m})$ and the acceleration $\Delta x_i \ddot{f}_i^{cm}(t_{m})$. 
After a time of flight $t_{t}$ the position of the cloud along $x$ is:
\begin{eqnarray}
x^{cm}_i(t_{m}+t_{t}) & = & \Delta x_i \left[ f_i^{cm}(t_{m}) + \dot{f}_i^{cm}(t_{m})t_{t} \right. \nonumber\\
& & \left. + \ddot{f}_i^{cm}(t_{m})\frac{t^2_{t}}{2} \right] \;.
\end{eqnarray}
The last equation shows that measuring $x^{cm}_i(t_{m}+t_{t})$ can be used to deduce $\Delta x_i$. 
Thus $\Delta x_i$ is used to scale the trajectories of the two states of the interferometer $x_i^{cm}(t) = \Delta x_i f_i^{cm} (t)$.


\section{Contrast decay versus $\Delta v$}
\label{Annexe_B_Contrast}

In appendix \ref{Annexe_A_ComputePos}, we have shown that the two outputs of the interferometer have different velocities,
which we will characterize as the difference $\Delta v = \dot{x}^{cm}_2 - \dot{x}^{cm}_1$.
This difference means that the paths are at least partially distinguishable and leads to a loss of interference contrast.
In section \ref{sec_SimpleContrast}, we first give an approximate model of the contrast decay versus $\Delta v$  in which we treat the spatial distribution at the output as a standing wave modulated by an envelope corresponding to the mean density.
The choice of the envelope is critical and we show that for the parameters of our experiment, there is a significant difference between using Boltzmann or Bose statistics for the atoms. 
We then proceed to describe a more accurate analytic model for the contrast decay.
In sections \ref{sec_Contrat_eq} and \ref{sec_Disgression}, we derive a general equation for the contrast $C$ as a function of $\Delta v$ which is applied to Boltzmann and Bose statistics in section \ref{sec_ExpressionContrast}.

While the approximate argument in section \ref{sec_SimpleContrast} takes the perspective of standing waves, sections \ref{sec_Contrat_eq}, \ref{sec_Disgression} and \ref{sec_ExpressionContrast} focus more on the issue of indistinguishability, using the fact that the wave functions in the harmonic trap have a velocity width.

\subsection{A simple model for the contrast decay versus $\Delta v$}
\label{sec_SimpleContrast}

Let us describe the two output states of the interferometer as two modulated plane waves:
\begin{equation}
\psi_i\left(x\right) = \sqrt{n_i\left(x\right)} \exp\left(j k_i x\right) \;,
\end{equation}
where $i$ is state $\left|1\right>$ or $\left|2\right>$, the wave vector is $k_i=m \dot{x}_i^{cm} /\hbar$ and $n_i\left(x\right)$ is the spatial density of atoms along $x$.
The interference of the two waves will lead to a modulation of the atomic density along $x$, which is proportional to:
\begin{eqnarray}
f(x) = \left| \psi_1 + \psi_2 \right|^2 = 2 n\left(x\right) \cos^2\left(\frac{m\Delta v}{2\hbar}x\right) \;,
\end{eqnarray}
for one internal state, and to:
\begin{eqnarray}
g(x) = \left| \psi_1 - \psi_2 \right|^2 = 2 n\left(x\right) \sin^2\left(\frac{m\Delta v}{2\hbar}x\right) \;,
\end{eqnarray}
for the other one. 
At the output of the interferometer, the two states overlap, thus $n\left(x\right)/2=n_1\left(x\right)=n_2\left(x\right)$.

We measure the number of atoms in each state irrespective of position, meaning that we integrate the functions $f(x)$ and $g(x)$ over $x$. 
The contrast $C\left(\Delta v\right)$ is given by:
\begin{equation}
C\left(\Delta v\right) = \left| \frac{\int_{-\infty}^{+\infty}f(x)dx-\int_{-\infty}^{+\infty}g(x)dx}{\int_{-\infty}^{+\infty}f(x)dx+\int_{-\infty}^{+\infty}g(x)dx} \right| \;.
\end{equation}

Assuming Boltzmann statistics, the atomic density along $x$, for a gas trapped in an harmonic potential of frequency $\omega_x/(2\pi)$, is:
\begin{equation}
n_{th}\left(x\right) = N \sqrt{\frac{m\omega_x^2}{2\pi k_B T}} \exp\left( - \frac{m\omega_x^2}{2k_B T}x^2 \right) \;,
\end{equation}
while for a Bose-Einstein statistics the density is:
\begin{eqnarray}
& & n_{BE}\left(x\right) = \frac{1}{N} \sqrt{\frac{mk_B T}{2\pi \hbar^2}} \nonumber\\
& \times & \sum_{l=1}^{+\infty} \frac{z^l}{l^{1/2}} \left[ \exp\left(- \frac{m\omega_x^2}{2k_B T}x^2 \right) \right]^l \;,
\end{eqnarray}
where $z$ is the gas fugacity, which, for $\hbar\omega_x \ll k_BT$, can be approximated as $z=N\hbar\omega_x/(k_BT)$.

\begin{figure}
\centering  \includegraphics[width=0.44\textwidth]{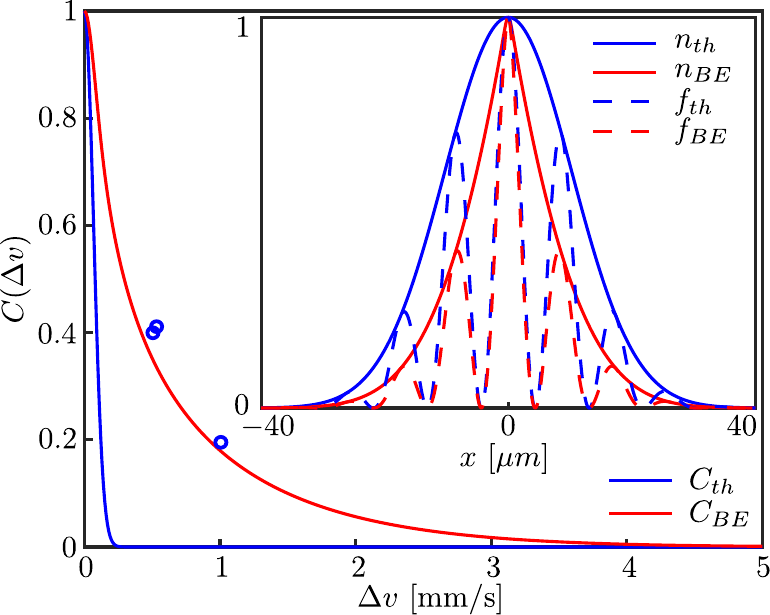}
\caption{\label{fig_ContrastSimple} (Color online) Approximated model of the contrast $C$ as a function of velocity difference between the two states at the output of the interferometer $\Delta v$ in mm/s, blue $C_{th}$ for a gas following a Boltzmann statistics, and red $C_{BE}$ following a Bose-Einstein statistics.
Solid lines in the inset show the atomic density along $x$ for a thermal gas $n_{th}$ (blue) and for a Bose-Einstein gas $n_{BE}$ (red).
Dashed lines in the inset show the density modulation along $x$ due to the interference at the output of the interferometer for $\Delta v = 0.5$~mm/s, $f_{th}$ in blue for a Boltzmann gas and $f_{BE}$ in red for Bose-Einstein gas.
All the curves are plotted for a temperature of $T=800$~nK and for an atom number $N=10^4$.
Blue dots are the experimental data of figure \ref{fig_Fringes}.}
\end{figure}

For Boltzmann statistics we find:
\begin{equation}
C_{th}\left(\Delta v\right) = \exp\left( -\frac{mk_B T}{2\hbar^2\omega_x^2}\left(\Delta v\right)^2 \right) \;.
\label{eq_CthSimple}
\end{equation}
This function is plotted in figure \ref{fig_ContrastSimple} for a temperature of $T=800$~nK. 
For Bose-Einstein statistics the contrast is:
\begin{eqnarray}
& & C_{BE}\left(\Delta v\right) = \frac{k_B T}{N \hbar \omega_x} \nonumber\\
& \times & \sum_{l=1}^{+\infty} \frac{z^l}{l}\exp\left( -\frac{mk_B T}{2l\hbar^2\omega_x^2}\left(\Delta v\right)^2 \right) \;,
\label{eq_CBESimple}
\end{eqnarray}
and is also plotted in figure \ref{fig_ContrastSimple}.
In the last expression, keeping only the term $l=1$, reduces $C_{BE}$ to $C_{th}$.

In the inset of figure \ref{fig_ContrastSimple}, we plotted the densities $n_{th}$ and $n_{BE}$ for a temperature of 800~nK and $10^4$ atoms, along with the resulting modulated standing wave for a typical value of $\Delta v$ in our experiment. 
It is seen that the density profiles are rather different and that this difference drastically changes the integral of the function $f(x)$.

\subsection{Definition of the contrast}
\label{sec_Contrat_eq}

In the following sections, we develop a more complete model for the contrast decay with $\Delta v$.
Before the first $\pi/2$ pulse of the interferometer the atom cloud is at rest and centered on the trap minimum. 
We denote by $p_n$ the population of each vibrational level.
All atoms are in the internal state $\left|2\right>$, thus the gas is described by the density matrix:
\begin{equation}
\hat{\rho}(0) = \sum_{n=0}^{+\infty} p_n \left|\phi_2^n(0)\right>\left|2\right> \left<2\right|\left<\phi_2^n(0)\right| \;.
\end{equation}
The state $\left|\phi_i^n(t)\right>\left|i\right>$ describes one atom in internal state $\left|i\right>$ trapped in the $n$th vibrational state $\left|\phi_i^n(t)\right>$ of the trap $V_{\left|i\right>}(t)$. 
Let us suppose that $V_{\left|1\right>}(0)=V_{\left|2\right>}(0)$ and $V_{\left|1\right>}(T_R)=V_{\left|2\right>}(T_R)$.
If the first $\pi/2$ pulse, is close to resonance and short compare to the Ramsey time and the inverse of the trap frequencies, it can be modeled as:
\begin{eqnarray}
\left|1\right> & \rightarrow & \frac{1}{\sqrt{2}}\left[ \left|1\right> - je^{+j\varphi}\left|2\right> \right] \;, \nonumber\\
\left|2\right> & \rightarrow & \frac{1}{\sqrt{2}}\left[ \left|2\right> - je^{-j\varphi}\left|1\right> \right] \;,
\end{eqnarray}
where $\varphi$ is the phase of the field driving the $\pi/2$ pulse at the begin of the pulse. 
After the first pulse $\left|\phi_2^n\right>\left|2\right>$ becomes:
\begin{equation}
\frac{1}{\sqrt{2}} \left[ \left|\phi_2^n(0)\right>\left|2\right> - je^{-j\varphi}\left|\phi_2^n(0)\right>\left|1\right> \right] \;.
\label{eq_StateAfterFirstPi2}
\end{equation}
During the Ramsey time $T_R$, the state $\left|\phi_2^n(t)\right>\left|2\right>$ follows the potential $V_{\left|2\right>}$ and stays in $\left|\phi_2^n(t)\right>\left|2\right>$. 
During the same time the state $\left|\phi_2^n(t)\right>\left|1\right>$ follows the potential $V_{\left|1\right>}$.
As $V_{\left|1\right>}(0)=V_{\left|2\right>}(0)$, we have $\left|\phi_2^n(0)\right>\left|1\right>=\left|\phi_1^n(0)\right>\left|1\right>$.
At the end of the Ramsey time, the state (\ref{eq_StateAfterFirstPi2}) is:
\begin{equation}
\frac{1}{\sqrt{2}} \left[ \left|\phi_2^n(T_R)\right>\left|2\right> - je^{-j\varphi}\left|\phi_1^n(T_R)\right>\left|1\right> \right] \;,
\label{eq_StateAfterTr}
\end{equation}
where the phase accumulated between the two states $\left|\phi_2^n(T_R)\right>\left|2\right>$ and $\left|\phi_1^n(T_R)\right>\left|1\right>$ during the Ramsey time $T_R$ is contained in the time dependent states $\left|\phi_i^n(t)\right>$.
Then the interferometer is closed with a second $\pi/2$ pulse, and the state (\ref{eq_StateAfterTr}) becomes:
\begin{eqnarray}
& & \left|\chi_n\right> = \frac{1}{2} \left[ \left|\phi_2^n(T_R)\right>\left|2\right> -je^{-j\varphi'}\left|\phi_2^n(T_R)\right>\left|1\right> - je^{-j\varphi}\times \right. \nonumber\\ 
& & \left. \left( -je^{+j\varphi'}\left|\phi_1^n(T_R)\right>\left|2\right> + \left|\phi_1^n(T_R)\right>\left|1\right> \right) \right] \;,
\end{eqnarray}
where $\varphi'$ is the phase of the field driving the $\pi/2$ pulse at the begining of the second pulse, thus if this field has a frequency $\omega_{osc}/(2\pi)$ we have $\omega_{osc} T_R = \varphi'-\varphi$. 
The Ramsey fringes on state $\left|1\right>$ (or $\left|2\right>$) are given by the total population $P_1$ (or $P_2$) in state $\left|1\right>$ (or $\left|2\right>$): 
\begin{eqnarray}
P_1 & = & \mathrm{Tr}\left( \rho(T_R) \left|1\right>\left<1\right| \right) \;, \nonumber\\
P_2 & = & \mathrm{Tr}\left( \rho(T_R) \left|2\right>\left<2\right| \right) \;,
\label{eq_PopDef}
\end{eqnarray}
with the density matrix of the gas at the output of the interferometer:
\begin{equation}
\hat{\rho}(T_R) = \sum_{n=0}^{+\infty} p_n \left|\chi_n\right> \left<\chi_n\right| \;,
\end{equation}
and the trace $\mathrm{Tr}$ is over the vibrational states $n$ and the spatial coordinate $x$.
Every computation done, using:
\begin{equation}
1 = \sum_{(k,n)=0}^{+\infty} p_n \left| \left<\phi_2^k(T_R) | \phi_1^n(T_R)\right> \right|^2 \;,
\end{equation}
and writing:
\begin{equation}
A(T_R) = \sum_{n=0}^{+\infty} p_n e^{j\omega T_R} \left<\phi_2^n(T_R) | \phi_1^n(T_R)\right> \;,
\label{eq_DefA}
\end{equation}
we found:
\begin{eqnarray}
P_1 & = & \frac{1}{2}\left[ 1 + \Re\left( A(T_R) \right) \right] \;, \\
P_2 & = & \frac{1}{2}\left[ 1 - \Re\left( A(T_R) \right) \right] \;,
\end{eqnarray}
where $\Re(z)$ stands for the real part of the complex number $z$. 
If, in equation (\ref{eq_PopDef}), we do not integrate over the spatial coordinate $x$, $P_1$ and $P_2$ are analogous to $f(x)$ and $g(x)$ of section \ref{sec_SimpleContrast}.
The contrast $C(T_R)$ and the phase shift $S(T_R)$ of those fringes are thus defined as:
\begin{equation}
C(T_R) = \left|A(T_R)\right| \;, \qquad S(T_R)=\arg\left(A(T_R)\right) \;.
\label{eq_DefConPhase}
\end{equation}

\subsection{Computation of the overlap}
\label{sec_Disgression}

\subsubsection{Expression of the wavefunction overlap}

To go further in the computation of the contrast, one needs to compute the spatial overlap of the vibrational states:
\begin{equation}
\left<\phi_2^k(T_R) | \phi_1^n(T_R)\right> = \int_{-\infty}^{+\infty} \phi_2^{k\dagger}(T_R,x) \phi_1^n(T_R,x) dx \;,
\end{equation}
and thus to know the $\phi_i^n(T_R,x)$ for the Hamiltonian acting the state $\left|i\right>$ during the Ramsey time:
\begin{eqnarray}
\hat{H}\left|i\right>\left<i\right| = \left[ \frac{p^2}{2m} + \frac{1}{2} m \omega_i^2(t) \left( x - x_{i}(t)  \right)^2 \right. \nonumber\\
\left. - \frac{1}{2}m\omega_i^2(t)x_{i}^2(t) \pm \frac{\hbar\Omega_{i}^2(t)}{4\Delta_i} \right] \left|i\right>\left<i\right| \;.
\end{eqnarray}
A generic solution of the Schrödinger equation with the above Hamiltonian is \cite{DupontNivet2014,DupontNivet2016,Husimi1953a,Husimi1953b,Popov1969,Popov1970,Lewis1969}:
\begin{equation}
\left|t\right>\left|i\right> = \sum_{n=0}^{+\infty} c_n^i\left|\phi_i^n(t)\right>\left|i\right> \;,
\end{equation}
where the $c_n^i$ are time-independent coefficients and $\phi_i^n(t,x)=\left< x | \phi_i^n(t) \right>$ are given by:
\begin{eqnarray}
& & \phi_i^n(t,x) = \left(\frac{m}{\pi\hbar}\right)^{1/4}\frac{\exp\left(j\alpha_n^i(t)\right)}{\sqrt{n!2^n\rho_i}} \nonumber\\
\times & \exp &\left(-\frac{m}{2\hbar}\left(\frac{x-x_i^{cm}}{\rho_i}\right)^2\right) H_n\left(\sqrt{\frac{m}{\hbar}}\frac{x-x_i^{cm}}{\rho_i}\right) \;. \nonumber\\
\end{eqnarray}
where $H_n$ are the Hermite polynomials, $\rho_i$ is solution of the equation:
\begin{equation}
\ddot{\rho}_i + \omega_i^2(t)\rho_i=\frac{1}{\rho_i^3} \;,
\end{equation}
and $\sqrt{m/(\hbar\rho_i^2)}$ is the size of the wavefunction.
The center of mass of the wavefunction $x^{cm}_i$ is a solution of the equation (\ref{eq_Xcm}).
The phases $\alpha_n^i(t)$ are given by
\begin{equation}
\alpha_n^i(t) = - \left(n+\frac{1}{2}\right)\int_0^t \frac{dt'}{\rho_i^2(t')} + u_i(x,t) + v_i(t) \;,
\end{equation}
with:
\begin{eqnarray}
u_i(x,t) & = & \frac{m}{\hbar}\left[ \frac{\dot{\rho}_i}{2\rho_i} + \frac{1}{\rho_i}\left(\dot{x}^{cm}_i\rho_i-x^{cm}_i\dot{\rho}_i\right)x \right] \;, \\
v_i(t) & = & \frac{m}{2\hbar}\int_0^t dt' \left[\frac{\left(\dot{x}^{cm}_i\rho_i-x^{cm}_i\dot{\rho}_i\right)^2}{\rho^2_i}\right] \nonumber\\
& & + \frac{m}{2\hbar}\int_0^t dt' \left[-\frac{\left(x^{cm}_i\right)^2}{\rho_i^4} + \omega_i^2x_{i}^2\right] \;.
\end{eqnarray}

Let us assume that during the Ramsey sequence the frequencies of both trapping potentials $V_{\left|1\right>}$ and $V_{\left|2\right>}$ are constant and are the same, thus $\omega_i(t) = \omega = 1/\rho_i^2$.
We will also assume that for the first and the second $\pi/2$ pulses, the states $\left|2\right>$ and $\left|1\right>$ are at the minimum of the trap: $x_2^{cm}(t=0)=x_1^{cm}(t=0)=0$ and $x_2^{cm}(t=T_R)=x_1^{cm}(t=T_R)=0$.
Thus:
\begin{eqnarray}
& & \alpha^2_n(T_R)-\alpha^1_n(T_R) = \frac{m}{\hbar}\left(\dot{x}_2^{cm}-\dot{x}_1^{cm}\right)x \nonumber\\
& & +\frac{m\omega^2}{2\hbar} \int_0^{T_R} dt'\left[ x_2^2(t')-\left(\dot{x}_2^{cm}\right)^2 - x_1^2(t')+\left(\dot{x}_1^{cm}\right)^2\right] \nonumber\\
& & +\frac{m}{2\hbar}\int_0^{T_R} dt'\left[ \left(\dot{x}_2^{cm}\right)^2 - \left(\dot{x}_1^{cm}\right)^2 \right] \;.
\end{eqnarray}
Then the spatial overlap $\left<\phi_2^n(T_R) | \phi_1^n(T_R)\right>$ of the wavefunctions is:
\begin{equation}
\left<\phi_2^n(T_R) | \phi_1^n(T_R)\right> = \frac{1}{2^nn!}\frac{e^{j\Phi_0}}{\sqrt{\pi}} \int dz e^{j\beta z} e^{-z^2}H_n^2(z) \;,
\end{equation}
$\Phi_0$ is a phase independent of $x$ and $n$ and thus will be discarded when computing the contrast. 
We have defined:
\begin{eqnarray}
\beta = \sqrt{\frac{m}{\hbar \omega}} \left(\dot{x}_2^{cm}-\dot{x}_1^{cm}\right) = \sqrt{\frac{m}{\hbar \omega}} \Delta v \;.
\end{eqnarray}

\subsubsection{Computation of the wavefunction overlap}

We now need to compute the following integral:
\begin{equation}
I_n(\beta) = \int_{-\infty}^{+\infty}dx H_n^2(x)e^{-x^2}e^{j\beta x} \;.
\end{equation}
We start with the generating function of the Hermite polynomials:
\begin{equation}
\exp\left( 2xt-t^2 \right) = \sum_{n=0}^{+\infty} H_n(x) \frac{t^n}{n!} \;,
\end{equation}
and apply the change of variable $t \leftarrow te^{j\theta}$:
\begin{equation}
\exp\left( 2xte^{j\theta} -t^2e^{2j\theta} \right) = \sum_{n=0}^{+\infty} H_n(x) e^{jn\theta} \frac{t^n}{n!} \;.
\end{equation}
We recognize the expansion of $f(\theta)=\exp\left( 2xte^{j\theta} -t^2e^{2j\theta} \right)$ in a Fourier series with the Fourier coefficients: $c_n=H_n(x)t^n/n!$. 
Using Parseval's equality:
\begin{equation}
\sum_n |c_n|^2 = \frac{1}{2\pi} \int_{-\pi}^{+\pi} |f(u)|^2 du \;,
\end{equation}
we find:
\begin{eqnarray}
& & 2\pi\sum_n H_n^2(x)\frac{t^{2n}}{n!^2} = \nonumber\\
& & \int_{-\pi}^{+\pi} \left| \exp\left( 2xte^{j\theta}-t^2e^{2j\theta} \right) \right|^2d\theta \;,
\end{eqnarray}
which can be rewritten as:
\begin{eqnarray}
& & 2\pi\sum_n H_n^2(x) \frac{t^{2n}}{n!^2}= \nonumber\\ 
& & \int_{-\pi}^{+\pi} \exp\left(4xt\cos\theta-2t^2\cos\left(2\theta\right)\right) d\theta \;.
\label{eq_eq1}
\end{eqnarray}
We multiply each side by $e^{-x^2+j\beta x}$ and integrate over $x$. 
For the left side we recognize the integral $I_n(\beta)$.
The right side of equation (\ref{eq_eq1}) becomes:
\begin{eqnarray}
& & \int_{-\infty}^{+\infty}dx\int_{-\pi}^{+\pi}d\theta \exp\left[ 4xt\cos\theta-2t^2\cos\left(2\theta\right) \right. \nonumber\\
& & \left. - x^2 +j\beta x \right] = \sqrt{\pi}\exp\left(-\frac{\beta^2}{4}\right) \nonumber\\
& & \qquad \times \int_{-\pi}^{+\pi}d\theta \exp\left(2t^2+2j\beta t\cos\theta\right) \;,
\end{eqnarray}

Introducing the Bessel function $J_0$:
\begin{equation}
J_0(x) = \frac{1}{2\pi}\int_{-\pi}^{+\pi} \exp\left(jx\cos\theta\right)d\theta \;,
\end{equation}
equation (\ref{eq_eq1}) can be recast as:
\begin{equation}
\sum_{n=0}^{+\infty} I_n(\beta)\frac{t^{2n}}{n!^2} = \sqrt{\pi}e^{-\frac{\beta^2}{4}+2t^2} J_0(2\beta t) \;.
\label{eq_eq2}
\end{equation}
$J_0$ and $e^{2t^2}$ can be written in power series:
\begin{eqnarray}
J_0(2\beta t) & = & \sum_{p=0}^{+\infty} \frac{(-1)^p}{p!^2} \left(\beta t\right)^{2p} \;, \\
e^{2t^2} & = & \sum_{n=0}^{+\infty} \frac{t^{2n}2^n}{n!} \;,
\end{eqnarray}
thus equation (\ref{eq_eq2}) becomes:
\begin{eqnarray}
& & \sum_{n=0}^{+\infty} \frac{t^{2n}}{n!^2} I_n(\beta) = \nonumber\\
& & \sqrt{\pi}e^{-\frac{\beta^2}{4}} \sum_{n=0}^{+\infty} \sum_{b=0}^n \frac{2^{n-b}(-1)^b \beta^{2b}}{(n-b)!b!^2} t^{2n} \;.
\label{eq_eq3}
\end{eqnarray}
We now introduce the Laguerre polynomials:
\begin{equation}
L_n(x) = \sum_{k=0}^n \frac{(-1)^kx^kn!}{k!^2(n-k)!} \;,
\end{equation}
and by identification of the coefficient of the polynomial in $t^{2n}$ in equation (\ref{eq_eq3}), we find:
\begin{equation}
I_n(\beta) = \sqrt{\pi} e^{-\frac{\beta^2}{4}} n! 2^n L_n\left(\frac{\beta^2}{2}\right) \;.
\label{eq_IntResult}
\end{equation}

\subsection{Expression of the contrast}
\label{sec_ExpressionContrast}

Replacing equation (\ref{eq_IntResult}) in equation (\ref{eq_DefA}) and using the contrast definition (\ref{eq_DefConPhase}), we have:
\begin{eqnarray}
C(\Delta v) & = & \left|\sum_{n=0}^{+\infty} p_n \left<\phi_2^n | \phi_1^n \right> \right| \nonumber\\
& = & \exp\left( -\frac{\beta^2}{4} \right) \left|\sum_{n=0}^{+\infty} p_n L_n\left(\frac{\beta^2}{2}\right) \right| \;.
\label{eq_ContrastGene}
\end{eqnarray}

\begin{figure}
\centering  \includegraphics[width=0.44\textwidth]{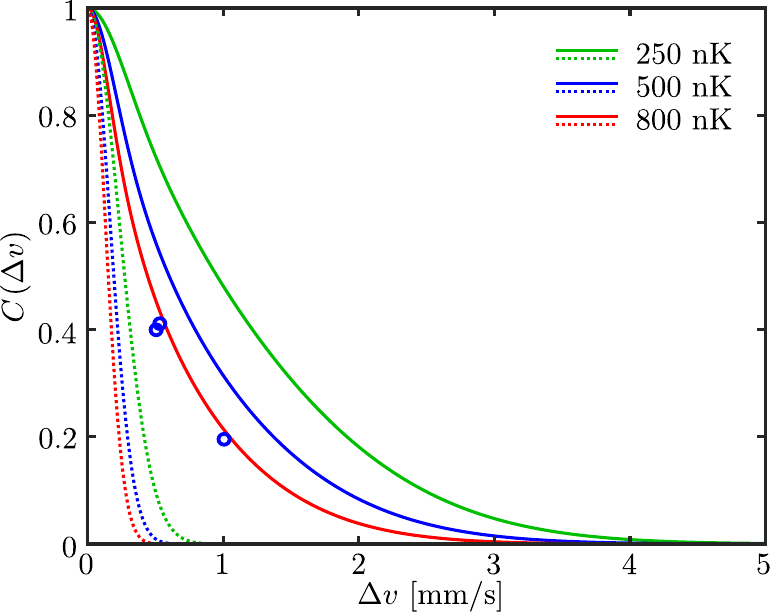}
\caption{\label{fig_Contrast} (Color online) Simulation of the contrast, $C$, as a function of velocity difference between the two states at the output of the interferometer $\Delta v$ in mm/s for different gas temperatures. 
Dashed lines show $C_{th}$ for a gas following a Boltzmann statistics (equation (\ref{eq_EqCth})), while solid lines show $C_{BE}$ following Bose-Einstein statistics (equation (\ref{eq_EqCBE})).
The parameters are $N=10^4$ atoms and $\overline{\omega}/(2\pi)=\left( 144\times322\times350 \right)^{1/3}$~Hz.
Blue dots are the experimental data of figure \ref{fig_Fringes} for a gas temperature of 800~nK.}
\end{figure}

\subsubsection{Case of a Boltzmann statistics}
\label{sec_ContrastTher}

Let us first consider the case of a Boltzmann statistics. The populations are given by:
\begin{equation}
p_n = \left(1-\lambda\right) \lambda^n \;, \qquad \lambda = \exp\left(-\frac{\hbar\overline{\omega}}{k_BT}\right) \;,
\end{equation}
where $\overline{\omega}$ is the geometric mean of the trap frequencies, $k_B$ the Boltzmann constant and $T$ the temperature. 
Using equation (\ref{eq_ContrastGene}) and the generating function of the Laguerre polynomials:
\begin{equation}
\sum_{n=0}^{+\infty} L_n(x) t^n = \frac{e^{-xt/(1-t)}}{1-t} \;,
\end{equation}
we find the following expression for the contrast:
\begin{equation}
C_{th}(\Delta v)=\exp\left( -\frac{\beta^2}{4}\frac{1+\lambda}{1-\lambda} \right) \;.
\label{eq_EqCth}
\end{equation}
This result is plotted in figure \ref{fig_Contrast}. If $\hbar\overline{\omega} \ll k_BT$, $\lambda$ can be approximated as $1 - \hbar\overline{\omega}/(k_BT)$, and the previous formula reduces to equation (\ref{eq_CthSimple}) derived in our simple contrast decay model.

\subsubsection{Case of a Bose-Einstein statistics}
\label{sec_ContrastBE}

If we consider a Bose-Einstein statistics in a one-dimensional trap, the $p_n$ are given by:
\begin{equation}
p_n=\frac{1}{N}\frac{z\lambda^n}{1-z\lambda^n} \;,
\end{equation}
where $N$ is the atom number in the gas and the fugacity $z$ is given by:
\begin{equation}
N = \sum_{n=0}^{+\infty} \frac{z\lambda^n}{1-z\lambda^n} \;,
\end{equation}
thus the contrast is given by:
\begin{eqnarray}
& & C_{BE}(\Delta v) = \frac{1}{N}\left| \sum_{n=0}^{+\infty} \frac{z\lambda^n}{1-z\lambda^n} L_n\left(\frac{\beta^2}{2}\right) \right| \nonumber\\
& = & \frac{1}{N}\left| \sum_{k=0}^{+\infty} z^{k+1} \sum_{n=0}^{+\infty} \left(\lambda^{k+1}\right)^n L_n\left(\frac{\beta^2}{2}\right) \right| \;.
\end{eqnarray}
Using the generating function of the Laguerre polynomials, we find the contrast in the case of a one-dimensional Bose gas:
\begin{eqnarray}
& & C_{BE}(\Delta v) = \nonumber\\
 \frac{1}{N}&&\left| \sum_{k=0}^{+\infty}\frac{z^{k+1}}{1-\lambda^{k+1}}\exp\left( -\frac{1+\lambda^{k+1}}{1-\lambda^{k+1}}\frac{\beta^2}{4} \right) \right| \;.
\label{eq_EqCBE}
\end{eqnarray}
The above equation is plotted in figure \ref{fig_Contrast}. If $\hbar\overline{\omega} (k+1) \ll k_BT$, $\lambda^{k+1}$ can be approximated as $1 - (k+1)\hbar\overline{\omega}/(k_BT)$, and the equation (\ref{eq_EqCBE}) reduces to equation (\ref{eq_CBESimple}) derived in our simple contrast decay model.


\section{Estimation of the trap symmetry}
\label{sec_TrapSym}

In appendix \ref{sec_TrapMin}, equation (\ref{eq_TrapFreqPos}) gives the dressed trap frequencies along $x$ from which we deduce, assuming that $\hbar\Omega_{i0}^2(\beta_i^2+2\alpha_i)/(2\Delta_i m)$ is small compared to $\omega$, the following expression:
\begin{eqnarray}
\omega_i=\omega\left( 1 \pm \frac{\hbar\Omega_{i0}^2(\beta_i^2+2\alpha_i)}{4\Delta_i m \omega^2} \right)  \;,
\end{eqnarray}
assuming also that the maxima of the two dressing field amplitudes are nearly equal $\Omega_{20}=\Omega_{10}+\delta\Omega$, with $\delta\Omega=2\pi\times$~300~Hz, as are their detunings $\Delta_{10}=-\Delta_{20}+\delta\Delta$, with $\delta\Delta=2\pi\times$~9~kHz, $\beta=|\beta_1|=|\beta_2|$, and $\alpha=\alpha_1=\alpha_2$, we find the trap symmetry:
\begin{eqnarray}
\frac{\delta\omega}{\omega} & = & \frac{\left|\omega_2-\omega_1\right|}{\omega} \nonumber\\
& = & \frac{\hbar\Omega_{10}^2(\beta^2+2\alpha)}{4\Delta_{10} m \omega^2}\left| \frac{2\delta\Omega}{\Omega_{10}}+\frac{\delta\Delta}{\Delta_{10}} \right| \;.
\end{eqnarray}
\\
The numbers in table \ref{tab_DressPara} give $\delta\omega/\omega\approx 1.5\cdot 10^{-3}$. References \cite{DupontNivet2014,DupontNivet2017b} link the contrast decay time to the trap symmetry:
\begin{equation}
t_c\approx \frac{1}{\delta\omega} \frac{\hbar\omega}{k_B T} \;,
\end{equation}
leading to $t_c\approx 6$~ms (for $T=800$~nK).
To simplify the computation of $t_c$ a time-independent model of $\delta\omega/\omega$ has been used leading to an overestimate of the trap symmetry. 
A time-dependent model can be found in reference \cite{DupontNivet2014}. 
The contrast decay law versus time $t$ is given in \cite{DupontNivet2017}:
\begin{equation}
C_{trap}\left(u=\frac{t}{t_c}\right) = \frac{1}{\left(u^2+1\right)^{3/2}} \;.
\end{equation}
For the above value of $t_c$ and for the time the microwave dressing fields are on, $t=1.9$~ms, we find a contrast of 87\%.

\bibliography{biblio}

\providecommand{\noopsort}[1]{}\providecommand{\singleletter}[1]{#1}%
\begin{thebibliography}{46}%
\makeatletter
\providecommand \@ifxundefined [1]{%
 \@ifx{#1\undefined}
}%
\providecommand \@ifnum [1]{%
 \ifnum #1\expandafter \@firstoftwo
 \else \expandafter \@secondoftwo
 \fi
}%
\providecommand \@ifx [1]{%
 \ifx #1\expandafter \@firstoftwo
 \else \expandafter \@secondoftwo
 \fi
}%
\providecommand \natexlab [1]{#1}%
\providecommand \enquote  [1]{``#1''}%
\providecommand \bibnamefont  [1]{#1}%
\providecommand \bibfnamefont [1]{#1}%
\providecommand \citenamefont [1]{#1}%
\providecommand \href@noop [0]{\@secondoftwo}%
\providecommand \href [0]{\begingroup \@sanitize@url \@href}%
\providecommand \@href[1]{\@@startlink{#1}\@@href}%
\providecommand \@@href[1]{\endgroup#1\@@endlink}%
\providecommand \@sanitize@url [0]{\catcode `\\12\catcode `\$12\catcode
  `\&12\catcode `\#12\catcode `\^12\catcode `\_12\catcode `\%12\relax}%
\providecommand \@@startlink[1]{}%
\providecommand \@@endlink[0]{}%
\providecommand \url  [0]{\begingroup\@sanitize@url \@url }%
\providecommand \@url [1]{\endgroup\@href {#1}{\urlprefix }}%
\providecommand \urlprefix  [0]{URL }%
\providecommand \Eprint [0]{\href }%
\providecommand \doibase [0]{http://dx.doi.org/}%
\providecommand \selectlanguage [0]{\@gobble}%
\providecommand \bibinfo  [0]{\@secondoftwo}%
\providecommand \bibfield  [0]{\@secondoftwo}%
\providecommand \translation [1]{[#1]}%
\providecommand \BibitemOpen [0]{}%
\providecommand \bibitemStop [0]{}%
\providecommand \bibitemNoStop [0]{.\EOS\space}%
\providecommand \EOS [0]{\spacefactor3000\relax}%
\providecommand \BibitemShut  [1]{\csname bibitem#1\endcsname}%
\let\auto@bib@innerbib\@empty
\bibitem [{\citenamefont {Salducci}\ \emph {et~al.}(2024)\citenamefont
  {Salducci}, \citenamefont {Bidel}, \citenamefont {Cadoret}, \citenamefont
  {Darmon}, \citenamefont {Zahzam}, \citenamefont {Bonnin}, \citenamefont
  {Schwartz}, \citenamefont {Blanchard},\ and\ \citenamefont
  {Bresson}}]{Salducci2024}%
  \BibitemOpen
  \bibfield  {author} {\bibinfo {author} {\bibfnamefont {C.}~\bibnamefont
  {Salducci}}, \bibinfo {author} {\bibfnamefont {Y.}~\bibnamefont {Bidel}},
  \bibinfo {author} {\bibfnamefont {M.}~\bibnamefont {Cadoret}}, \bibinfo
  {author} {\bibfnamefont {S.}~\bibnamefont {Darmon}}, \bibinfo {author}
  {\bibfnamefont {N.}~\bibnamefont {Zahzam}}, \bibinfo {author} {\bibfnamefont
  {A.}~\bibnamefont {Bonnin}}, \bibinfo {author} {\bibfnamefont
  {S.}~\bibnamefont {Schwartz}}, \bibinfo {author} {\bibfnamefont
  {C.}~\bibnamefont {Blanchard}}, \ and\ \bibinfo {author} {\bibfnamefont
  {A.}~\bibnamefont {Bresson}},\ }\href
  {https://doi.org/10.1126/sciadv.adq4498} {\bibfield  {journal} {\bibinfo
  {journal} {Sci. Adv.}\ }\textbf {\bibinfo {volume} {10}},\ \bibinfo {pages}
  {eadq4498} (\bibinfo {year} {2024})}\BibitemShut {NoStop}%
\bibitem [{\citenamefont {Abend}\ \emph {et~al.}(2016)\citenamefont {Abend},
  \citenamefont {Gebbe}, \citenamefont {Gersemann}, \citenamefont {Ahlers},
  \citenamefont {M{\"u}ntinga}, \citenamefont {Giese}, \citenamefont {Gaaloul},
  \citenamefont {Schubert}, \citenamefont {L{\"a}mmerzahl}, \citenamefont
  {Ertmer}, \citenamefont {Schleich},\ and\ \citenamefont {Rasel}}]{Abend2016}%
  \BibitemOpen
  \bibfield  {author} {\bibinfo {author} {\bibfnamefont {S.}~\bibnamefont
  {Abend}}, \bibinfo {author} {\bibfnamefont {M.}~\bibnamefont {Gebbe}},
  \bibinfo {author} {\bibfnamefont {M.}~\bibnamefont {Gersemann}}, \bibinfo
  {author} {\bibfnamefont {H.}~\bibnamefont {Ahlers}}, \bibinfo {author}
  {\bibfnamefont {H.}~\bibnamefont {M{\"u}ntinga}}, \bibinfo {author}
  {\bibfnamefont {E.}~\bibnamefont {Giese}}, \bibinfo {author} {\bibfnamefont
  {N.}~\bibnamefont {Gaaloul}}, \bibinfo {author} {\bibfnamefont
  {C.}~\bibnamefont {Schubert}}, \bibinfo {author} {\bibfnamefont
  {C.}~\bibnamefont {L{\"a}mmerzahl}}, \bibinfo {author} {\bibfnamefont
  {W.}~\bibnamefont {Ertmer}}, \bibinfo {author} {\bibfnamefont {W.~P.}\
  \bibnamefont {Schleich}}, \ and\ \bibinfo {author} {\bibfnamefont {E.~M.}\
  \bibnamefont {Rasel}},\ }\href
  {https://doi.org/10.1103/PhysRevLett.117.203003} {\bibfield  {journal}
  {\bibinfo  {journal} {Phys. Rev. Lett.}\ }\textbf {\bibinfo {volume} {117}},\
  \bibinfo {pages} {203003} (\bibinfo {year} {2016})}\BibitemShut {NoStop}%
\bibitem [{\citenamefont {Savoie}\ \emph {et~al.}(2018)\citenamefont {Savoie},
  \citenamefont {Altorio}, \citenamefont {Fang}, \citenamefont {Sidorenkov},
  \citenamefont {Geiger},\ and\ \citenamefont {Landragin}}]{Savoie2018}%
  \BibitemOpen
  \bibfield  {author} {\bibinfo {author} {\bibfnamefont {D.}~\bibnamefont
  {Savoie}}, \bibinfo {author} {\bibfnamefont {M.}~\bibnamefont {Altorio}},
  \bibinfo {author} {\bibfnamefont {B.}~\bibnamefont {Fang}}, \bibinfo {author}
  {\bibfnamefont {L.}~\bibnamefont {Sidorenkov}}, \bibinfo {author}
  {\bibfnamefont {R.}~\bibnamefont {Geiger}}, \ and\ \bibinfo {author}
  {\bibfnamefont {A.}~\bibnamefont {Landragin}},\ }\href
  {https://doi.org/10.1126/sciadv.aau7948} {\bibfield  {journal} {\bibinfo
  {journal} {Sci. Adv.}\ }\textbf {\bibinfo {volume} {4}},\ \bibinfo {pages}
  {eaau7948} (\bibinfo {year} {2018})}\BibitemShut {NoStop}%
\bibitem [{\citenamefont {Cronin}\ \emph {et~al.}(2009)\citenamefont {Cronin},
  \citenamefont {Schmiedmayer},\ and\ \citenamefont {Pritchard}}]{Cronin2009}%
  \BibitemOpen
  \bibfield  {author} {\bibinfo {author} {\bibfnamefont {A.}~\bibnamefont
  {Cronin}}, \bibinfo {author} {\bibfnamefont {J.}~\bibnamefont
  {Schmiedmayer}}, \ and\ \bibinfo {author} {\bibfnamefont {D.}~\bibnamefont
  {Pritchard}},\ }\href {\doibase 10.1103/RevModPhys.81.1051} {\bibfield
  {journal} {\bibinfo  {journal} {Rev. Mod. Phys.}\ }\textbf {\bibinfo {volume}
  {81}},\ \bibinfo {pages} {1051} (\bibinfo {year} {2009})}\BibitemShut
  {NoStop}%
\bibitem [{\citenamefont {Bongs}\ \emph {et~al.}(2019)\citenamefont {Bongs},
  \citenamefont {Holynski}, \citenamefont {Vovrosh}, \citenamefont {Bouyer},
  \citenamefont {Condon}, \citenamefont {Rasel}, \citenamefont {Schubert},
  \citenamefont {Schleich},\ and\ \citenamefont {Roura}}]{Bongs2019}%
  \BibitemOpen
  \bibfield  {author} {\bibinfo {author} {\bibfnamefont {K.}~\bibnamefont
  {Bongs}}, \bibinfo {author} {\bibfnamefont {M.}~\bibnamefont {Holynski}},
  \bibinfo {author} {\bibfnamefont {J.}~\bibnamefont {Vovrosh}}, \bibinfo
  {author} {\bibfnamefont {P.}~\bibnamefont {Bouyer}}, \bibinfo {author}
  {\bibfnamefont {G.}~\bibnamefont {Condon}}, \bibinfo {author} {\bibfnamefont
  {E.}~\bibnamefont {Rasel}}, \bibinfo {author} {\bibfnamefont
  {C.}~\bibnamefont {Schubert}}, \bibinfo {author} {\bibfnamefont {W.~P.}\
  \bibnamefont {Schleich}}, \ and\ \bibinfo {author} {\bibfnamefont
  {A.}~\bibnamefont {Roura}},\ }\href
  {https://doi.org/10.1038/s42254-019-0117-4} {\bibfield  {journal} {\bibinfo
  {journal} {Nat. Rev. Phys.}\ }\textbf {\bibinfo {volume} {1}},\ \bibinfo
  {pages} {731} (\bibinfo {year} {2019})}\BibitemShut {NoStop}%
\bibitem [{\citenamefont {Geiger}\ \emph {et~al.}(2020)\citenamefont {Geiger},
  \citenamefont {Landragin}, \citenamefont {Merlet},\ and\ \citenamefont
  {Pereira Dos~Santos}}]{Geiger2020}%
  \BibitemOpen
  \bibfield  {author} {\bibinfo {author} {\bibfnamefont {R.}~\bibnamefont
  {Geiger}}, \bibinfo {author} {\bibfnamefont {A.}~\bibnamefont {Landragin}},
  \bibinfo {author} {\bibfnamefont {S.}~\bibnamefont {Merlet}}, \ and\ \bibinfo
  {author} {\bibfnamefont {F.}~\bibnamefont {Pereira Dos~Santos}},\ }\href
  {https://doi.org/10.1116/5.0009093} {\bibfield  {journal} {\bibinfo
  {journal} {AVS Quantum Sci.}\ }\textbf {\bibinfo {volume} {2}} (\bibinfo
  {year} {2020})}\BibitemShut {NoStop}%
\bibitem [{\citenamefont {Ammar}\ \emph {et~al.}(2015)\citenamefont {Ammar},
  \citenamefont {Dupont-Nivet}, \citenamefont {Huet}, \citenamefont {Pocholle},
  \citenamefont {Rosenbusch}, \citenamefont {Bouchoule}, \citenamefont
  {Westbrook}, \citenamefont {Est\`eve}, \citenamefont {Reichel}, \citenamefont
  {Guerlin},\ and\ \citenamefont {Schwartz}}]{Ammar2014}%
  \BibitemOpen
  \bibfield  {author} {\bibinfo {author} {\bibfnamefont {M.}~\bibnamefont
  {Ammar}}, \bibinfo {author} {\bibfnamefont {M.}~\bibnamefont {Dupont-Nivet}},
  \bibinfo {author} {\bibfnamefont {L.}~\bibnamefont {Huet}}, \bibinfo {author}
  {\bibfnamefont {J.-P.}\ \bibnamefont {Pocholle}}, \bibinfo {author}
  {\bibfnamefont {P.}~\bibnamefont {Rosenbusch}}, \bibinfo {author}
  {\bibfnamefont {I.}~\bibnamefont {Bouchoule}}, \bibinfo {author}
  {\bibfnamefont {C.~I.}\ \bibnamefont {Westbrook}}, \bibinfo {author}
  {\bibfnamefont {J.}~\bibnamefont {Est\`eve}}, \bibinfo {author}
  {\bibfnamefont {J.}~\bibnamefont {Reichel}}, \bibinfo {author} {\bibfnamefont
  {C.}~\bibnamefont {Guerlin}}, \ and\ \bibinfo {author} {\bibfnamefont
  {S.}~\bibnamefont {Schwartz}},\ }\href {\doibase 10.1103/PhysRevA.91.053623}
  {\bibfield  {journal} {\bibinfo  {journal} {Phys. Rev. A}\ }\textbf {\bibinfo
  {volume} {91}},\ \bibinfo {pages} {053623} (\bibinfo {year}
  {2015})}\BibitemShut {NoStop}%
\bibitem [{\citenamefont {B{\"o}hi}\ \emph {et~al.}(2009)\citenamefont
  {B{\"o}hi}, \citenamefont {Riedel}, \citenamefont {Hoffrogge}, \citenamefont
  {Reichel}, \citenamefont {Hansch},\ and\ \citenamefont
  {Treutlein}}]{Bohi2009}%
  \BibitemOpen
  \bibfield  {author} {\bibinfo {author} {\bibfnamefont {P.}~\bibnamefont
  {B{\"o}hi}}, \bibinfo {author} {\bibfnamefont {M.}~\bibnamefont {Riedel}},
  \bibinfo {author} {\bibfnamefont {J.}~\bibnamefont {Hoffrogge}}, \bibinfo
  {author} {\bibfnamefont {J.}~\bibnamefont {Reichel}}, \bibinfo {author}
  {\bibfnamefont {T.}~\bibnamefont {Hansch}}, \ and\ \bibinfo {author}
  {\bibfnamefont {P.}~\bibnamefont {Treutlein}},\ }\href
  {https://doi.org/10.1038/nphys1329} {\bibfield  {journal} {\bibinfo
  {journal} {Nat. Phys.}\ }\textbf {\bibinfo {volume} {5}},\ \bibinfo {pages}
  {592} (\bibinfo {year} {2009})}\BibitemShut {NoStop}%
\bibitem [{\citenamefont {Schumm}\ \emph {et~al.}(2005)\citenamefont {Schumm},
  \citenamefont {Hofferberth}, \citenamefont {Andersson}, \citenamefont
  {Wildermuth}, \citenamefont {Groth}, \citenamefont {Bar-Joseph},
  \citenamefont {Schmiedmayer},\ and\ \citenamefont {Kruger}}]{Schumm2005}%
  \BibitemOpen
  \bibfield  {author} {\bibinfo {author} {\bibfnamefont {T.}~\bibnamefont
  {Schumm}}, \bibinfo {author} {\bibfnamefont {S.}~\bibnamefont {Hofferberth}},
  \bibinfo {author} {\bibfnamefont {L.~M.}\ \bibnamefont {Andersson}}, \bibinfo
  {author} {\bibfnamefont {S.}~\bibnamefont {Wildermuth}}, \bibinfo {author}
  {\bibfnamefont {S.}~\bibnamefont {Groth}}, \bibinfo {author} {\bibfnamefont
  {I.}~\bibnamefont {Bar-Joseph}}, \bibinfo {author} {\bibfnamefont
  {J.}~\bibnamefont {Schmiedmayer}}, \ and\ \bibinfo {author} {\bibfnamefont
  {P.}~\bibnamefont {Kruger}},\ }\href {https://doi.org/10.1038/nphys125}
  {\bibfield  {journal} {\bibinfo  {journal} {Nat. Phys.}\ }\textbf {\bibinfo
  {volume} {1}},\ \bibinfo {pages} {57} (\bibinfo {year} {2005})}\BibitemShut
  {NoStop}%
\bibitem [{\citenamefont {Schumm}\ \emph {et~al.}(2006)\citenamefont {Schumm},
  \citenamefont {Kr{\"u}ger}, \citenamefont {Hofferberth}, \citenamefont
  {Lesanovsky}, \citenamefont {Wildermuth}, \citenamefont {Groth},
  \citenamefont {Bar-Joseph}, \citenamefont {Andersson},\ and\ \citenamefont
  {Schmiedmayer}}]{Schumm2006}%
  \BibitemOpen
  \bibfield  {author} {\bibinfo {author} {\bibfnamefont {T.}~\bibnamefont
  {Schumm}}, \bibinfo {author} {\bibfnamefont {P.}~\bibnamefont {Kr{\"u}ger}},
  \bibinfo {author} {\bibfnamefont {S.}~\bibnamefont {Hofferberth}}, \bibinfo
  {author} {\bibfnamefont {I.}~\bibnamefont {Lesanovsky}}, \bibinfo {author}
  {\bibfnamefont {S.}~\bibnamefont {Wildermuth}}, \bibinfo {author}
  {\bibfnamefont {S.}~\bibnamefont {Groth}}, \bibinfo {author} {\bibfnamefont
  {I.}~\bibnamefont {Bar-Joseph}}, \bibinfo {author} {\bibfnamefont {L.~M.}\
  \bibnamefont {Andersson}}, \ and\ \bibinfo {author} {\bibfnamefont
  {J.}~\bibnamefont {Schmiedmayer}},\ }\href
  {https://doi.org/10.1007/s11128-006-0033-2} {\bibfield  {journal} {\bibinfo
  {journal} {Quantum Inf. Process.}\ }\textbf {\bibinfo {volume} {5}},\
  \bibinfo {pages} {537} (\bibinfo {year} {2006})}\BibitemShut {NoStop}%
\bibitem [{\citenamefont {Wang}\ \emph {et~al.}(2005)\citenamefont {Wang},
  \citenamefont {Anderson}, \citenamefont {Bright}, \citenamefont {Cornell},
  \citenamefont {Diot}, \citenamefont {Kishimoto}, \citenamefont {Prentiss},
  \citenamefont {Saravanan}, \citenamefont {Segal},\ and\ \citenamefont
  {Wu}}]{Wang2005}%
  \BibitemOpen
  \bibfield  {author} {\bibinfo {author} {\bibfnamefont {Y.-J.}\ \bibnamefont
  {Wang}}, \bibinfo {author} {\bibfnamefont {D.~Z.}\ \bibnamefont {Anderson}},
  \bibinfo {author} {\bibfnamefont {V.~M.}\ \bibnamefont {Bright}}, \bibinfo
  {author} {\bibfnamefont {E.~A.}\ \bibnamefont {Cornell}}, \bibinfo {author}
  {\bibfnamefont {Q.}~\bibnamefont {Diot}}, \bibinfo {author} {\bibfnamefont
  {T.}~\bibnamefont {Kishimoto}}, \bibinfo {author} {\bibfnamefont
  {M.}~\bibnamefont {Prentiss}}, \bibinfo {author} {\bibfnamefont {R.~A.}\
  \bibnamefont {Saravanan}}, \bibinfo {author} {\bibfnamefont {S.~R.}\
  \bibnamefont {Segal}}, \ and\ \bibinfo {author} {\bibfnamefont
  {S.}~\bibnamefont {Wu}},\ }\href {\doibase 10.1103/PhysRevLett.94.090405}
  {\bibfield  {journal} {\bibinfo  {journal} {Phys. Rev. Lett.}\ }\textbf
  {\bibinfo {volume} {94}},\ \bibinfo {pages} {090405} (\bibinfo {year}
  {2005})}\BibitemShut {NoStop}%
\bibitem [{\citenamefont {Petrovic}\ \emph {et~al.}(2013)\citenamefont
  {Petrovic}, \citenamefont {Herrera}, \citenamefont {Lombardi}, \citenamefont
  {Schäfer},\ and\ \citenamefont {Cataliotti}}]{Petrovic2013}%
  \BibitemOpen
  \bibfield  {author} {\bibinfo {author} {\bibfnamefont {J.}~\bibnamefont
  {Petrovic}}, \bibinfo {author} {\bibfnamefont {I.}~\bibnamefont {Herrera}},
  \bibinfo {author} {\bibfnamefont {P.}~\bibnamefont {Lombardi}}, \bibinfo
  {author} {\bibfnamefont {F.}~\bibnamefont {Schäfer}}, \ and\ \bibinfo
  {author} {\bibfnamefont {F.~S.}\ \bibnamefont {Cataliotti}},\ }\href
  {http://stacks.iop.org/1367-2630/15/i=4/a=043002} {\bibfield  {journal}
  {\bibinfo  {journal} {New J. Phys.}\ }\textbf {\bibinfo {volume} {15}},\
  \bibinfo {pages} {043002} (\bibinfo {year} {2013})}\BibitemShut {NoStop}%
\bibitem [{\citenamefont {Fancher}\ \emph {et~al.}(2018)\citenamefont
  {Fancher}, \citenamefont {Pyle}, \citenamefont {Rotunno},\ and\ \citenamefont
  {Aubin}}]{Fancher2018}%
  \BibitemOpen
  \bibfield  {author} {\bibinfo {author} {\bibfnamefont {C.~T.}\ \bibnamefont
  {Fancher}}, \bibinfo {author} {\bibfnamefont {A.~J.}\ \bibnamefont {Pyle}},
  \bibinfo {author} {\bibfnamefont {A.~P.}\ \bibnamefont {Rotunno}}, \ and\
  \bibinfo {author} {\bibfnamefont {S.}~\bibnamefont {Aubin}},\ }\href
  {\doibase 10.1103/PhysRevA.97.043430} {\bibfield  {journal} {\bibinfo
  {journal} {Phys. Rev. A}\ }\textbf {\bibinfo {volume} {97}},\ \bibinfo
  {pages} {043430} (\bibinfo {year} {2018})}\BibitemShut {NoStop}%
\bibitem [{\citenamefont {Riedel}\ \emph {et~al.}(2010)\citenamefont {Riedel},
  \citenamefont {B{\"o}hi}, \citenamefont {Li}, \citenamefont {H{\"a}nsch},
  \citenamefont {Sinatra},\ and\ \citenamefont {Treutlein}}]{Riedel2010}%
  \BibitemOpen
  \bibfield  {author} {\bibinfo {author} {\bibfnamefont {M.~F.}\ \bibnamefont
  {Riedel}}, \bibinfo {author} {\bibfnamefont {P.}~\bibnamefont {B{\"o}hi}},
  \bibinfo {author} {\bibfnamefont {Y.}~\bibnamefont {Li}}, \bibinfo {author}
  {\bibfnamefont {T.~W.}\ \bibnamefont {H{\"a}nsch}}, \bibinfo {author}
  {\bibfnamefont {A.}~\bibnamefont {Sinatra}}, \ and\ \bibinfo {author}
  {\bibfnamefont {P.}~\bibnamefont {Treutlein}},\ }\href
  {https://doi.org/10.1038/nature08988} {\bibfield  {journal} {\bibinfo
  {journal} {Nature}\ }\textbf {\bibinfo {volume} {464}},\ \bibinfo {pages}
  {1170} (\bibinfo {year} {2010})}\BibitemShut {NoStop}%
\bibitem [{\citenamefont {Agosta}\ \emph {et~al.}(1989)\citenamefont {Agosta},
  \citenamefont {Silvera}, \citenamefont {Stoof},\ and\ \citenamefont
  {Verhaar}}]{Agosta1989}%
  \BibitemOpen
  \bibfield  {author} {\bibinfo {author} {\bibfnamefont {C.~C.}\ \bibnamefont
  {Agosta}}, \bibinfo {author} {\bibfnamefont {I.~F.}\ \bibnamefont {Silvera}},
  \bibinfo {author} {\bibfnamefont {H.~T.~C.}\ \bibnamefont {Stoof}}, \ and\
  \bibinfo {author} {\bibfnamefont {B.~J.}\ \bibnamefont {Verhaar}},\ }\href
  {\doibase 10.1103/PhysRevLett.62.2361} {\bibfield  {journal} {\bibinfo
  {journal} {Phys. Rev. Lett.}\ }\textbf {\bibinfo {volume} {62}},\ \bibinfo
  {pages} {2361} (\bibinfo {year} {1989})}\BibitemShut {NoStop}%
\bibitem [{\citenamefont {Perrin}\ and\ \citenamefont
  {Garraway}(2017)}]{Perrin2017}%
  \BibitemOpen
  \bibfield  {author} {\bibinfo {author} {\bibfnamefont {H.}~\bibnamefont
  {Perrin}}\ and\ \bibinfo {author} {\bibfnamefont {B.~M.}\ \bibnamefont
  {Garraway}},\ }\bibfield  {booktitle} {\emph {\bibinfo {booktitle} {Advances
  in Atomic, Molecular, and Optical Physics}},\ }\href
  {https://doi.org/10.1016/bs.aamop.2017.03.002} {\ \textbf {\bibinfo {volume}
  {66}},\ \bibinfo {pages} {181} (\bibinfo {year} {2017})}\BibitemShut
  {NoStop}%
\bibitem [{\citenamefont {Dupont-Nivet}\ \emph {et~al.}(2016)\citenamefont
  {Dupont-Nivet}, \citenamefont {Westbrook},\ and\ \citenamefont
  {Schwartz}}]{DupontNivet2014}%
  \BibitemOpen
  \bibfield  {author} {\bibinfo {author} {\bibfnamefont {M.}~\bibnamefont
  {Dupont-Nivet}}, \bibinfo {author} {\bibfnamefont {C.~I.}\ \bibnamefont
  {Westbrook}}, \ and\ \bibinfo {author} {\bibfnamefont {S.}~\bibnamefont
  {Schwartz}},\ }\href {http://dx.doi.org/10.1088/1367-2630/18/11/113012}
  {\bibfield  {journal} {\bibinfo  {journal} {New J. Phys.}\ }\textbf {\bibinfo
  {volume} {18}},\ \bibinfo {pages} {113012} (\bibinfo {year}
  {2016})}\BibitemShut {NoStop}%
\bibitem [{\citenamefont {Dupont-Nivet}\ \emph {et~al.}(2018)\citenamefont
  {Dupont-Nivet}, \citenamefont {Demur}, \citenamefont {Westbrook},\ and\
  \citenamefont {Schwartz}}]{DupontNivet2017b}%
  \BibitemOpen
  \bibfield  {author} {\bibinfo {author} {\bibfnamefont {M.}~\bibnamefont
  {Dupont-Nivet}}, \bibinfo {author} {\bibfnamefont {R.}~\bibnamefont {Demur}},
  \bibinfo {author} {\bibfnamefont {C.~I.}\ \bibnamefont {Westbrook}}, \ and\
  \bibinfo {author} {\bibfnamefont {S.}~\bibnamefont {Schwartz}},\ }\href
  {https://doi.org/10.1088/1367-2630/aabc72} {\bibfield  {journal} {\bibinfo
  {journal} {New J. Phys.}\ }\textbf {\bibinfo {volume} {20}},\ \bibinfo
  {pages} {043051} (\bibinfo {year} {2018})}\BibitemShut {NoStop}%
\bibitem [{\citenamefont {Harber}\ \emph {et~al.}(2002)\citenamefont {Harber},
  \citenamefont {Lewandowski}, \citenamefont {McGuirk},\ and\ \citenamefont
  {Cornell}}]{Harber2002}%
  \BibitemOpen
  \bibfield  {author} {\bibinfo {author} {\bibfnamefont {D.~M.}\ \bibnamefont
  {Harber}}, \bibinfo {author} {\bibfnamefont {H.~J.}\ \bibnamefont
  {Lewandowski}}, \bibinfo {author} {\bibfnamefont {J.~M.}\ \bibnamefont
  {McGuirk}}, \ and\ \bibinfo {author} {\bibfnamefont {E.~A.}\ \bibnamefont
  {Cornell}},\ }\href {\doibase 10.1103/PhysRevA.66.053616} {\bibfield
  {journal} {\bibinfo  {journal} {Phys. Rev. A}\ }\textbf {\bibinfo {volume}
  {66}},\ \bibinfo {pages} {053616} (\bibinfo {year} {2002})}\BibitemShut
  {NoStop}%
\bibitem [{\citenamefont {Treutlein}\ \emph {et~al.}(2004)\citenamefont
  {Treutlein}, \citenamefont {Hommelhoff}, \citenamefont {Steinmetz},
  \citenamefont {H\"ansch},\ and\ \citenamefont {Reichel}}]{Treutlein2004}%
  \BibitemOpen
  \bibfield  {author} {\bibinfo {author} {\bibfnamefont {P.}~\bibnamefont
  {Treutlein}}, \bibinfo {author} {\bibfnamefont {P.}~\bibnamefont
  {Hommelhoff}}, \bibinfo {author} {\bibfnamefont {T.}~\bibnamefont
  {Steinmetz}}, \bibinfo {author} {\bibfnamefont {T.~W.}\ \bibnamefont
  {H\"ansch}}, \ and\ \bibinfo {author} {\bibfnamefont {J.}~\bibnamefont
  {Reichel}},\ }\href {\doibase 10.1103/PhysRevLett.92.203005} {\bibfield
  {journal} {\bibinfo  {journal} {Phys. Rev. Lett.}\ }\textbf {\bibinfo
  {volume} {92}},\ \bibinfo {pages} {203005} (\bibinfo {year}
  {2004})}\BibitemShut {NoStop}%
\bibitem [{\citenamefont {Szmuk}\ \emph {et~al.}(2015)\citenamefont {Szmuk},
  \citenamefont {Dugrain}, \citenamefont {Maineult}, \citenamefont {Reichel},\
  and\ \citenamefont {Rosenbusch}}]{Szmuk2015}%
  \BibitemOpen
  \bibfield  {author} {\bibinfo {author} {\bibfnamefont {R.}~\bibnamefont
  {Szmuk}}, \bibinfo {author} {\bibfnamefont {V.}~\bibnamefont {Dugrain}},
  \bibinfo {author} {\bibfnamefont {W.}~\bibnamefont {Maineult}}, \bibinfo
  {author} {\bibfnamefont {J.}~\bibnamefont {Reichel}}, \ and\ \bibinfo
  {author} {\bibfnamefont {P.}~\bibnamefont {Rosenbusch}},\ }\href
  {https://doi.org/10.1103/PhysRevA.92.012106} {\bibfield  {journal} {\bibinfo
  {journal} {Phys. Rev. A}\ }\textbf {\bibinfo {volume} {92}},\ \bibinfo
  {pages} {012106} (\bibinfo {year} {2015})}\BibitemShut {NoStop}%
\bibitem [{\citenamefont {Dupont-Nivet}\ \emph {et~al.}(2025)\citenamefont
  {Dupont-Nivet}, \citenamefont {Wirtschafter}, \citenamefont {Hello},\ and\
  \citenamefont {Westbrook}}]{DupontNivet2025}%
  \BibitemOpen
  \bibfield  {author} {\bibinfo {author} {\bibfnamefont {M.}~\bibnamefont
  {Dupont-Nivet}}, \bibinfo {author} {\bibfnamefont {B.}~\bibnamefont
  {Wirtschafter}}, \bibinfo {author} {\bibfnamefont {S.}~\bibnamefont {Hello}},
  \ and\ \bibinfo {author} {\bibfnamefont {C.~I.}\ \bibnamefont {Westbrook}},\
  }\href {\doibase 10.1103/PhysRevA.111.063106} {\bibfield  {journal} {\bibinfo
   {journal} {Phys. Rev. A}\ }\textbf {\bibinfo {volume} {111}},\ \bibinfo
  {pages} {063106} (\bibinfo {year} {2025})}\BibitemShut {NoStop}%
\bibitem [{\citenamefont {Huet}\ \emph {et~al.}(2012)\citenamefont {Huet},
  \citenamefont {Ammar}, \citenamefont {Morvan}, \citenamefont {Sarazin},
  \citenamefont {Pocholle}, \citenamefont {Reichel}, \citenamefont {Guerlin},\
  and\ \citenamefont {Schwartz}}]{Huet2012}%
  \BibitemOpen
  \bibfield  {author} {\bibinfo {author} {\bibfnamefont {L.}~\bibnamefont
  {Huet}}, \bibinfo {author} {\bibfnamefont {M.}~\bibnamefont {Ammar}},
  \bibinfo {author} {\bibfnamefont {E.}~\bibnamefont {Morvan}}, \bibinfo
  {author} {\bibfnamefont {N.}~\bibnamefont {Sarazin}}, \bibinfo {author}
  {\bibfnamefont {J.-P.}\ \bibnamefont {Pocholle}}, \bibinfo {author}
  {\bibfnamefont {J.}~\bibnamefont {Reichel}}, \bibinfo {author} {\bibfnamefont
  {C.}~\bibnamefont {Guerlin}}, \ and\ \bibinfo {author} {\bibfnamefont
  {S.}~\bibnamefont {Schwartz}},\ }\href {\doibase
  http://dx.doi.org/10.1063/1.3689777} {\bibfield  {journal} {\bibinfo
  {journal} {Appl. Phys. Lett.}\ }\textbf {\bibinfo {volume} {100}},\ \bibinfo
  {eid} {121114} (\bibinfo {year} {2012})}\BibitemShut {NoStop}%
\bibitem [{\citenamefont {Huet}(2013)}]{Huet2013}%
  \BibitemOpen
  \bibfield  {author} {\bibinfo {author} {\bibfnamefont {L.}~\bibnamefont
  {Huet}},\ }\emph {\bibinfo {title} {Gravim{\'e}trie atomique sur puce et
  applications embarqu{\'e}es}},\ \href
  {https://theses.hal.science/tel-00839785/} {Ph.D. thesis},\ \bibinfo
  {school} {Universit{\'e} Paris-Est} (\bibinfo {year} {2013})\BibitemShut
  {NoStop}%
\bibitem [{\citenamefont {Dupont-Nivet}(2016)}]{DupontNivet2016}%
  \BibitemOpen
  \bibfield  {author} {\bibinfo {author} {\bibfnamefont {M.}~\bibnamefont
  {Dupont-Nivet}},\ }\emph {\bibinfo {title} {Vers un acc\'el\'erom\'etre
  atomique sur puce}},\ \href {https://pastel.hal.science/tel-01366681/} {Ph.D.
  thesis},\ \bibinfo  {school} {Universit\'e Paris Saclay} (\bibinfo {year}
  {2016})\BibitemShut {NoStop}%
\bibitem [{\citenamefont {Wirtschafter}(2022)}]{Wirtschafter2022}%
  \BibitemOpen
  \bibfield  {author} {\bibinfo {author} {\bibfnamefont {B.}~\bibnamefont
  {Wirtschafter}},\ }\emph {\bibinfo {title} {Interferom\`etre à atomes froids
  pi\'eg\'es sur puce avec s\'eparation spatiale}},\ \href
  {https://pastel.hal.science/tel-04190518/} {Ph.D. thesis},\ \bibinfo
  {school} {Universit\'e Paris Saclay} (\bibinfo {year} {2022})\BibitemShut
  {NoStop}%
\bibitem [{\citenamefont {Farkas}\ \emph {et~al.}(2010)\citenamefont {Farkas},
  \citenamefont {Hudek}, \citenamefont {Salim}, \citenamefont {Segal},
  \citenamefont {Squires},\ and\ \citenamefont {Anderson}}]{Farkas2010}%
  \BibitemOpen
  \bibfield  {author} {\bibinfo {author} {\bibfnamefont {D.}~\bibnamefont
  {Farkas}}, \bibinfo {author} {\bibfnamefont {K.}~\bibnamefont {Hudek}},
  \bibinfo {author} {\bibfnamefont {E.}~\bibnamefont {Salim}}, \bibinfo
  {author} {\bibfnamefont {S.}~\bibnamefont {Segal}}, \bibinfo {author}
  {\bibfnamefont {M.}~\bibnamefont {Squires}}, \ and\ \bibinfo {author}
  {\bibfnamefont {D.}~\bibnamefont {Anderson}},\ }\href {\doibase
  http://dx.doi.org/10.1063/1.3327812} {\bibfield  {journal} {\bibinfo
  {journal} {Appl. Phys. Lett.}\ }\textbf {\bibinfo {volume} {96}},\ \bibinfo
  {eid} {093102} (\bibinfo {year} {2010})}\BibitemShut {NoStop}%
\bibitem [{\citenamefont {Squires}(2008)}]{Squires2008}%
  \BibitemOpen
  \bibfield  {author} {\bibinfo {author} {\bibfnamefont {M.~B.}\ \bibnamefont
  {Squires}},\ }\emph {\bibinfo {title} {High repetition rate Bose-Einstein
  condensate production in a compact, transportable vacuum system}},\ \href
  {https://citeseerx.ist.psu.edu/document?repid=rep1&type=pdf&doi=9a4f80c50ea1cf1329343d8073801e3e61673231}
  {Ph.D. thesis} (\bibinfo {year} {2008})\BibitemShut {NoStop}%
\bibitem [{\citenamefont {Dieckmann}\ \emph {et~al.}(1998)\citenamefont
  {Dieckmann}, \citenamefont {Spreeuw}, \citenamefont {Weidem\"uller},\ and\
  \citenamefont {Walraven}}]{Dieckmann1998}%
  \BibitemOpen
  \bibfield  {author} {\bibinfo {author} {\bibfnamefont {K.}~\bibnamefont
  {Dieckmann}}, \bibinfo {author} {\bibfnamefont {R.~J.~C.}\ \bibnamefont
  {Spreeuw}}, \bibinfo {author} {\bibfnamefont {M.}~\bibnamefont
  {Weidem\"uller}}, \ and\ \bibinfo {author} {\bibfnamefont {J.~T.~M.}\
  \bibnamefont {Walraven}},\ }\href {\doibase 10.1103/PhysRevA.58.3891}
  {\bibfield  {journal} {\bibinfo  {journal} {Phys. Rev. A}\ }\textbf {\bibinfo
  {volume} {58}},\ \bibinfo {pages} {3891} (\bibinfo {year}
  {1998})}\BibitemShut {NoStop}%
\bibitem [{\citenamefont {Schoser}\ \emph {et~al.}(2002)\citenamefont
  {Schoser}, \citenamefont {Bat\"ar}, \citenamefont {L\"ow}, \citenamefont
  {Schweikhard}, \citenamefont {Grabowski}, \citenamefont {Ovchinnikov},\ and\
  \citenamefont {Pfau}}]{Schoser2002}%
  \BibitemOpen
  \bibfield  {author} {\bibinfo {author} {\bibfnamefont {J.}~\bibnamefont
  {Schoser}}, \bibinfo {author} {\bibfnamefont {A.}~\bibnamefont {Bat\"ar}},
  \bibinfo {author} {\bibfnamefont {R.}~\bibnamefont {L\"ow}}, \bibinfo
  {author} {\bibfnamefont {V.}~\bibnamefont {Schweikhard}}, \bibinfo {author}
  {\bibfnamefont {A.}~\bibnamefont {Grabowski}}, \bibinfo {author}
  {\bibfnamefont {Y.~B.}\ \bibnamefont {Ovchinnikov}}, \ and\ \bibinfo {author}
  {\bibfnamefont {T.}~\bibnamefont {Pfau}},\ }\href {\doibase
  10.1103/PhysRevA.66.023410} {\bibfield  {journal} {\bibinfo  {journal} {Phys.
  Rev. A}\ }\textbf {\bibinfo {volume} {66}},\ \bibinfo {pages} {023410}
  (\bibinfo {year} {2002})}\BibitemShut {NoStop}%
\bibitem [{\citenamefont {Dupont-Nivet}\ \emph {et~al.}(2015)\citenamefont
  {Dupont-Nivet}, \citenamefont {Casiulis}, \citenamefont {Laudat},
  \citenamefont {Westbrook},\ and\ \citenamefont {Schwartz}}]{DupontNivet2015}%
  \BibitemOpen
  \bibfield  {author} {\bibinfo {author} {\bibfnamefont {M.}~\bibnamefont
  {Dupont-Nivet}}, \bibinfo {author} {\bibfnamefont {M.}~\bibnamefont
  {Casiulis}}, \bibinfo {author} {\bibfnamefont {T.}~\bibnamefont {Laudat}},
  \bibinfo {author} {\bibfnamefont {C.~I.}\ \bibnamefont {Westbrook}}, \ and\
  \bibinfo {author} {\bibfnamefont {S.}~\bibnamefont {Schwartz}},\ }\href
  {\doibase 10.1103/PhysRevA.91.053420} {\bibfield  {journal} {\bibinfo
  {journal} {Phys. Rev. A}\ }\textbf {\bibinfo {volume} {91}},\ \bibinfo
  {pages} {053420} (\bibinfo {year} {2015})}\BibitemShut {NoStop}%
\bibitem [{\citenamefont {Vitanov}\ \emph {et~al.}(2017)\citenamefont
  {Vitanov}, \citenamefont {Rangelov}, \citenamefont {Shore},\ and\
  \citenamefont {Bergmann}}]{Vitanov2017}%
  \BibitemOpen
  \bibfield  {author} {\bibinfo {author} {\bibfnamefont {N.~V.}\ \bibnamefont
  {Vitanov}}, \bibinfo {author} {\bibfnamefont {A.~A.}\ \bibnamefont
  {Rangelov}}, \bibinfo {author} {\bibfnamefont {B.~W.}\ \bibnamefont {Shore}},
  \ and\ \bibinfo {author} {\bibfnamefont {K.}~\bibnamefont {Bergmann}},\
  }\href {https://doi.org/10.1103/RevModPhys.89.015006} {\bibfield  {journal}
  {\bibinfo  {journal} {Rev. Mod. Phys.}\ }\textbf {\bibinfo {volume} {89}},\
  \bibinfo {pages} {015006} (\bibinfo {year} {2017})}\BibitemShut {NoStop}%
\bibitem [{\citenamefont {Amri}\ \emph {et~al.}(2019)\citenamefont {Amri},
  \citenamefont {Corgier}, \citenamefont {Sugny}, \citenamefont {Rasel},
  \citenamefont {Gaaloul},\ and\ \citenamefont {Charron}}]{Amri2019}%
  \BibitemOpen
  \bibfield  {author} {\bibinfo {author} {\bibfnamefont {S.}~\bibnamefont
  {Amri}}, \bibinfo {author} {\bibfnamefont {R.}~\bibnamefont {Corgier}},
  \bibinfo {author} {\bibfnamefont {D.}~\bibnamefont {Sugny}}, \bibinfo
  {author} {\bibfnamefont {E.~M.}\ \bibnamefont {Rasel}}, \bibinfo {author}
  {\bibfnamefont {N.}~\bibnamefont {Gaaloul}}, \ and\ \bibinfo {author}
  {\bibfnamefont {E.}~\bibnamefont {Charron}},\ }\href
  {https://doi.org/10.1038/s41598-019-41784-z} {\bibfield  {journal} {\bibinfo
  {journal} {Sci. Rep.}\ }\textbf {\bibinfo {volume} {9}},\ \bibinfo {pages}
  {5346} (\bibinfo {year} {2019})}\BibitemShut {NoStop}%
\bibitem [{\citenamefont {Corgier}\ \emph {et~al.}(2018)\citenamefont
  {Corgier}, \citenamefont {Amri}, \citenamefont {Herr}, \citenamefont
  {Ahlers}, \citenamefont {Rudolph}, \citenamefont {Gu{\'e}ry-Odelin},
  \citenamefont {Rasel}, \citenamefont {Charron},\ and\ \citenamefont
  {Gaaloul}}]{Corgier2018}%
  \BibitemOpen
  \bibfield  {author} {\bibinfo {author} {\bibfnamefont {R.}~\bibnamefont
  {Corgier}}, \bibinfo {author} {\bibfnamefont {S.}~\bibnamefont {Amri}},
  \bibinfo {author} {\bibfnamefont {W.}~\bibnamefont {Herr}}, \bibinfo {author}
  {\bibfnamefont {H.}~\bibnamefont {Ahlers}}, \bibinfo {author} {\bibfnamefont
  {J.}~\bibnamefont {Rudolph}}, \bibinfo {author} {\bibfnamefont
  {D.}~\bibnamefont {Gu{\'e}ry-Odelin}}, \bibinfo {author} {\bibfnamefont
  {E.~M.}\ \bibnamefont {Rasel}}, \bibinfo {author} {\bibfnamefont
  {E.}~\bibnamefont {Charron}}, \ and\ \bibinfo {author} {\bibfnamefont
  {N.}~\bibnamefont {Gaaloul}},\ }\href
  {https://doi.org/10.1088/1367-2630/aabdfc} {\bibfield  {journal} {\bibinfo
  {journal} {New J. Phys.}\ }\textbf {\bibinfo {volume} {20}},\ \bibinfo
  {pages} {055002} (\bibinfo {year} {2018})}\BibitemShut {NoStop}%
\bibitem [{\citenamefont {Ness}\ \emph {et~al.}(2018)\citenamefont {Ness},
  \citenamefont {Shkedrov}, \citenamefont {Florshaim},\ and\ \citenamefont
  {Sagi}}]{Ness2018}%
  \BibitemOpen
  \bibfield  {author} {\bibinfo {author} {\bibfnamefont {G.}~\bibnamefont
  {Ness}}, \bibinfo {author} {\bibfnamefont {C.}~\bibnamefont {Shkedrov}},
  \bibinfo {author} {\bibfnamefont {Y.}~\bibnamefont {Florshaim}}, \ and\
  \bibinfo {author} {\bibfnamefont {Y.}~\bibnamefont {Sagi}},\ }\href
  {https://doi.org/10.1088/1367-2630/aadcc1} {\bibfield  {journal} {\bibinfo
  {journal} {New J. Phys.}\ }\textbf {\bibinfo {volume} {20}},\ \bibinfo
  {pages} {095002} (\bibinfo {year} {2018})}\BibitemShut {NoStop}%
\bibitem [{\citenamefont {Hello}(2025)}]{Hello2025c}%
  \BibitemOpen
  \bibfield  {author} {\bibinfo {author} {\bibfnamefont {S.}~\bibnamefont
  {Hello}},\ }\emph {\bibinfo {title} {Capteur inertiel \`a atomes froids
  pi\'eg\'es sur puce}},\ \href@noop {} {Ph.D. thesis},\ \bibinfo  {school}
  {Universit\'e Paris Saclay} (\bibinfo {year} {2025})\BibitemShut {NoStop}%
\bibitem [{\citenamefont {Hello}\ \emph {et~al.}(2025)\citenamefont {Hello},
  \citenamefont {Mersch}, \citenamefont {Wirtschafter}, \citenamefont
  {Seguineau}, \citenamefont {Westbrook},\ and\ \citenamefont
  {Dupont-Nivet}}]{Hello2025b}%
  \BibitemOpen
  \bibfield  {author} {\bibinfo {author} {\bibfnamefont {S.}~\bibnamefont
  {Hello}}, \bibinfo {author} {\bibfnamefont {A.}~\bibnamefont {Mersch}},
  \bibinfo {author} {\bibfnamefont {B.}~\bibnamefont {Wirtschafter}}, \bibinfo
  {author} {\bibfnamefont {F.}~\bibnamefont {Seguineau}}, \bibinfo {author}
  {\bibfnamefont {C.~I.}\ \bibnamefont {Westbrook}}, \ and\ \bibinfo {author}
  {\bibfnamefont {M.}~\bibnamefont {Dupont-Nivet}},\ }in\ \href
  {https://doi.org/10.1109/INERTIAL63280.2025.11036959} {\emph {\bibinfo
  {booktitle} {2025 IEEE International Symposium on Inertial Sensors and
  Systems (INERTIAL)}}}\ (\bibinfo {organization} {IEEE},\ \bibinfo {year}
  {2025})\ pp.\ \bibinfo {pages} {1--4}\BibitemShut {NoStop}%
\bibitem [{\citenamefont {Wadell}(1991)}]{Wadell1991}%
  \BibitemOpen
  \bibfield  {author} {\bibinfo {author} {\bibfnamefont {B.~C.}\ \bibnamefont
  {Wadell}},\ }\href@noop {} {\emph {\bibinfo {title} {Transmission line design
  handbook}}}\ (\bibinfo  {publisher} {Artech House Publishers},\ \bibinfo
  {year} {1991})\BibitemShut {NoStop}%
\bibitem [{\citenamefont {Ammar}(2014)}]{Ammar2014b}%
  \BibitemOpen
  \bibfield  {author} {\bibinfo {author} {\bibfnamefont {M.}~\bibnamefont
  {Ammar}},\ }\emph {\bibinfo {title} {Design and study of microwave potentials
  for interferometry with thermal atoms on a chip}},\ \href
  {https://theses.hal.science/tel-01134188/} {Ph.D. thesis},\ \bibinfo
  {school} {Université Pierre et Marie Curie} (\bibinfo {year}
  {2014})\BibitemShut {NoStop}%
\bibitem [{\citenamefont {H.~R.~Lewis}\ and\ \citenamefont
  {Riesenfeld}(1969)}]{Lewis1969}%
  \BibitemOpen
  \bibfield  {author} {\bibinfo {author} {\bibfnamefont {J.}~\bibnamefont
  {H.~R.~Lewis}}\ and\ \bibinfo {author} {\bibfnamefont {W.~B.}\ \bibnamefont
  {Riesenfeld}},\ }\href {\doibase 10.1063/1.1664991} {\bibfield  {journal}
  {\bibinfo  {journal} {J. Math. Phys.}\ }\textbf {\bibinfo {volume} {10}},\
  \bibinfo {pages} {1458} (\bibinfo {year} {1969})}\BibitemShut {NoStop}%
\bibitem [{\citenamefont {Schaff}\ \emph {et~al.}(2011)\citenamefont {Schaff},
  \citenamefont {Capuzzi}, \citenamefont {Labeyrie},\ and\ \citenamefont
  {Vignolo}}]{Schaff2011}%
  \BibitemOpen
  \bibfield  {author} {\bibinfo {author} {\bibfnamefont {J.-F.}\ \bibnamefont
  {Schaff}}, \bibinfo {author} {\bibfnamefont {P.}~\bibnamefont {Capuzzi}},
  \bibinfo {author} {\bibfnamefont {G.}~\bibnamefont {Labeyrie}}, \ and\
  \bibinfo {author} {\bibfnamefont {P.}~\bibnamefont {Vignolo}},\ }\href
  {http://stacks.iop.org/1367-2630/13/i=11/a=113017} {\bibfield  {journal}
  {\bibinfo  {journal} {New J. Phys.}\ }\textbf {\bibinfo {volume} {13}},\
  \bibinfo {pages} {113017} (\bibinfo {year} {2011})}\BibitemShut {NoStop}%
\bibitem [{\citenamefont {Husimi}(1953{\natexlab{a}})}]{Husimi1953a}%
  \BibitemOpen
  \bibfield  {author} {\bibinfo {author} {\bibfnamefont {K.}~\bibnamefont
  {Husimi}},\ }\href {https://doi.org/10.1143/ptp/9.3.238} {\bibfield
  {journal} {\bibinfo  {journal} {Prog. Theor. Phys.}\ }\textbf {\bibinfo
  {volume} {9}},\ \bibinfo {pages} {238} (\bibinfo {year}
  {1953}{\natexlab{a}})}\BibitemShut {NoStop}%
\bibitem [{\citenamefont {Husimi}(1953{\natexlab{b}})}]{Husimi1953b}%
  \BibitemOpen
  \bibfield  {author} {\bibinfo {author} {\bibfnamefont {K.}~\bibnamefont
  {Husimi}},\ }\href {https://doi.org/10.1143/ptp/9.4.381} {\bibfield
  {journal} {\bibinfo  {journal} {Prog. Theor. Phys.}\ }\textbf {\bibinfo
  {volume} {9}},\ \bibinfo {pages} {381} (\bibinfo {year}
  {1953}{\natexlab{b}})}\BibitemShut {NoStop}%
\bibitem [{\citenamefont {Popov}\ and\ \citenamefont
  {Perelomov}(1969)}]{Popov1969}%
  \BibitemOpen
  \bibfield  {author} {\bibinfo {author} {\bibfnamefont {V.~S.}\ \bibnamefont
  {Popov}}\ and\ \bibinfo {author} {\bibfnamefont {A.~M.}\ \bibnamefont
  {Perelomov}},\ }\href@noop {} {\bibfield  {journal} {\bibinfo  {journal} {J.
  Exp. Theor. Phys.}\ }\textbf {\bibinfo {volume} {29}},\ \bibinfo {pages}
  {738} (\bibinfo {year} {1969})}\BibitemShut {NoStop}%
\bibitem [{\citenamefont {Popov}\ and\ \citenamefont
  {Perelomov}(1970)}]{Popov1970}%
  \BibitemOpen
  \bibfield  {author} {\bibinfo {author} {\bibfnamefont {V.~S.}\ \bibnamefont
  {Popov}}\ and\ \bibinfo {author} {\bibfnamefont {A.~M.}\ \bibnamefont
  {Perelomov}},\ }\href@noop {} {\bibfield  {journal} {\bibinfo  {journal} {J.
  Exp. Theor. Phys.}\ }\textbf {\bibinfo {volume} {30}},\ \bibinfo {pages}
  {910} (\bibinfo {year} {1970})}\BibitemShut {NoStop}%
\bibitem [{\citenamefont {Dupont-Nivet}\ \emph {et~al.}(2021)\citenamefont
  {Dupont-Nivet}, \citenamefont {Westbrook},\ and\ \citenamefont
  {Schwartz}}]{DupontNivet2017}%
  \BibitemOpen
  \bibfield  {author} {\bibinfo {author} {\bibfnamefont {M.}~\bibnamefont
  {Dupont-Nivet}}, \bibinfo {author} {\bibfnamefont {C.~I.}\ \bibnamefont
  {Westbrook}}, \ and\ \bibinfo {author} {\bibfnamefont {S.}~\bibnamefont
  {Schwartz}},\ }\href {https://doi.org/10.1103/PhysRevA.103.023321} {\bibfield
   {journal} {\bibinfo  {journal} {Phys. Rev. A}\ }\textbf {\bibinfo {volume}
  {103}},\ \bibinfo {pages} {023321} (\bibinfo {year} {2021})}\BibitemShut
  {NoStop}%
\end{thebibliography}%

\end{document}